\documentclass[a4paper,fleqn,usenatbib]{mnras}

\usepackage[T1]{fontenc}
\usepackage{ae,aecompl}

\usepackage{graphicx}	% Including figure files
\usepackage{amsmath}	% Advanced maths commands
\usepackage{amssymb}	% Extra maths symbols
\hypersetup
{
    pdfauthor={Kokubo and Minezaki},
    pdftitle={Kokubo and Minezaki 2019},
}

\title[Rapid Variations of the Dust Inner Radius in Mrk~590]{Rapid Luminosity Decline and Subsequent Reformation of the Innermost Dust Distribution in the Changing-look AGN Mrk 590}

\author[M.~Kokubo \& T.~Minezaki]{
Mitsuru Kokubo,$^{1}$\thanks{E-mail: mkokubo@astr.tohoku.ac.jp}\thanks{JSPS fellow}
Takeo Minezaki,$^{2}$
\\
$^{1}$ Astronomical Institute, Tohoku University, 6-3 Aramak-Aza-Aoba, Aoba-ku, Sendai, Miyagi 980-8578, Japan\\
$^{2}$ Institute of Astronomy, the University of Tokyo, 2-21-1 Osawa, Mitaka, Tokyo 181-0015, Japan\\
}

\date{Accepted XXX. Received YYY; in original form ZZZ}

\pubyear{2019}

\begin{document}
\label{firstpage}
\pagerange{\pageref{firstpage}--\pageref{lastpage}}
\maketitle

\begin{abstract}

We examine the long-term optical/near-infrared (NIR) flux variability of a ``changing-look'' active galactic nucleus (AGN) Mrk~590 between 1998 and 2007.
Multi-band multi-epoch optical/NIR photometry data from the SDSS Stripe 82 database and the Multicolor Active Galactic Nuclei Monitoring (MAGNUM) project reveal that Mrk~590 experienced a sudden luminosity decrease during the period from 2000 to 2001.
Detection of dust reverberation lag signals between $V$- and $K$-band light curves obtained by the MAGNUM project during the faint state in $2003-2007$ suggests that the dust torus innermost radius $R_\text{dust}$ of Mrk~590 had become very small [$R_\text{dust} \simeq 32$~ light-days (lt-days)] by the year 2004 according to the aforementioned significant decrease in AGN luminosity.
The $R_\text{dust}$ in the faint state is comparable to the H$\beta$ broad line region (BLR) radius of $R_{\text{H}\beta, \text{BLR}} \simeq 26$~lt-days measured by previous reverberation mapping observations during the bright state of Mrk~590 in $1990-1996$.
These observations indicate that the innermost radius of the dust torus in Mrk~590 decreased rapidly after the AGN ultraviolet-optical luminosity drop, and that the replenishment time scale of the innermost dust distribution is less than 4 years, which is much shorter than the free fall time scale of BLR gas or dust clouds.
We suggest that rapid replenishment of the innermost dust distribution can be accomplished either by new dust formation in radiatively-cooled BLR gas clouds or by new dust formation in the disk atmosphere and subsequent vertical wind from the dusty disk as a result of radiation pressure.

\end{abstract}

\begin{keywords}
accretion, accretion discs --
dust, extinction --
galaxies: active --
quasars: general  --
quasars: individual (Mrk~590)
\end{keywords}

%%%%%%%%%%%%%%%%%%%%%%%%%%%%%%%%%%%%%%%%%%%%%%%%%%

%%%%%%%%%%%%%%%%% BODY OF PAPER %%%%%%%%%%%%%%%%%%

%-------------------------------------------------------------------

\section{Introduction}
\label{sec:intro}

Active galactic nuclei (AGNs) are generally characterised by ultraviolet (UV)-optical blue continua from accretion disks surrounding supermassive black holes (SMBHs), broad emission lines (BELs), and hot dust emission at near-infrared (NIR) wavelengths.
The BELs from AGNs are produced in a gaseous region, called the broad line region (BLR), by reprocessing the incident ionizing radiation from the inner regions of the accretion disk  \citep[e.g.,][]{ost86}.
The innermost part of the dust torus surrounding the disk/BLR region is responsible for the IR thermal emission from the hot dust.
The dust grains in the dust torus are in radiative equilibrium with the disk UV radiation, and the innermost radius of the dust torus corresponds to the dust sublimation radius where the equilibrium temperature of the dust grains equals the dust sublimation temperature of $\sim$~$1,400-2,000$~K \citep[e.g.,][]{bar87,lao93,bas18}.
The AGN UV-optical accretion disk emission generally shows time variability on time scales of days to years \citep[e.g.,][]{kel09,mac10,mac12}; thus, the BELs and the dust IR continuum inevitably change their luminosities with time delays relative to disk continuum variations, which is commonly assumed to correspond to light travel times from the disk to the BLR and dust torus innermost radius ($R_{\text{dust}}$), respectively \citep[see, e.g.,][for detailed modeling of the BEL/dust IR continuum response to the disk continuum variations]{pan14,gar17,alm17}.

The BLR (dust) reverberation mapping method is used to measure the time lag between the UV-optical disk continuum light curve and the BEL (IR) light curve of an AGN and, consequently, to estimate the size of the otherwise spatially unresolved BLR radius (dust torus innermost radius) in the AGN \citep[][and references therein]{pet04b,sug06,kos14}.
There have been several BLR and dust reverberation mapping observation campaigns to date.
The BLR reverberation mapping observations for local Seyfert galaxies and some high redshift quasars have been carried out by several groups \citep[e.g.,][]{pet98b,pet04,kas00,ben09b,den10,gri12,du14,fau17b}; a compilation of $\sim 40$ AGNs with secure H$\beta$ BLR lag detection can be found in \cite{ben15} \citep[see also][]{ben09,ben13}.
Dust reverberation mapping observations have also been measured for several AGNs \citep{cla89,gla92,sit93,gla04,sug06,kos14,poz14,poz15,sch15,lir15,vaz15,man18,ram18,lan19}.
The first systematic investigation of dust reverberation mapping was carried out by the Multicolour Active Galactic Nuclei Monitoring (MAGNUM) project \citep[][]{yos02,yos03,min04,tom06,sug06,kos09,sak10,kos14,yos14}.
The MAGNUM project presents accurate $K$-band dust reverberation lag measurements for 17 local Seyfert~1 galaxies \citep{sug06,kos14}.

Among the 17 AGNs with dust reverberation lag measurements obtained by MAGNUM \citep{kos14}, 15 have literature values for the H$\beta$ BLR lag (see Section~\ref{blr_dust_radii} for more details).
Comparisons between the H$\beta$ BLR and dust reverberation lags in a sample of AGNs reveal that the dust innermost radii are generally larger compared to the H$\beta$ BLR radii, by a factor of $\sim 4$ \citep{sug06,kos14,du15,ben16}.
However, it has also been pointed out by several authors that the relationship between the BLR and the dust innermost radii ($R_{\text{H}\beta~\text{BLR}}$ and $R_\text{dust}$, respectively) in some AGNs seems to deviate from the general trend of $R_\text{dust} = 4 \times R_{\text{H}\beta~\text{BLR}}$.
As recapitulated in Section~\ref{blr_dust_radii}, Mrk~590 is the most significant outlier in the BLR$-$dust radius relation of the AGNs \citep{gan15,du15}; the dust radius of Mrk~590 is only slightly larger than the H$\beta$ radius (by a factor of $\sim 1.4$).

Mrk~590 is known as a ``changing-look'' AGN \citep[e.g.,][]{lam15}.
Two spectroscopic measurements of the AGN optical continuum emission obtained in 1996 and 2003 indicate that Mrk~590 experienced a large luminosity decrease, by a factor of $\sim 30$, during this period, and further luminosity declines were observed in 2006 and 2013 \citep{den14}.
The BEL luminosities became weaker associated with this decrease in the AGN luminosity, and \cite{den14} conclude that the BELs had disappeared from the optical spectrum obtained in 2013; thus, Mrk~590 had changed from an optical classification as Seyfert~1 to Seyfert~$1.9-2$ by 2013 \citep[see also][]{riv12,koa16a,koa16b,rai19}.
MAGNUM dust reverberation mapping observations for Mrk~590 were carried out in 2003$-$2007, when Mrk~590 was in a faint state.
However, H$\beta$ BLR reverberation mapping monitoring for Mrk~590 was carried out in 1990-1996 \citep{pet98b,pet04,ben09,ben13}, when Mrk~590 was in the brighter state.

Considering the different luminosity states between the epochs of the BLR and the dust reverberation monitoring observations, we hypothesise that the relatively small $R_\text{dust}$-to-$R_{\text{H}\beta~\text{BLR}}$ ratio in Mrk~590 is a result of intrinsic variation in the dust innermost radius due to the significant UV-optical luminosity decrease that occurred between the mid-1990s and the early 2000s.
The main purpose of this work is to verify this hypothesis quantitatively by analysing the temporal evolution of the AGN optical luminosity and dust innermost radius of Mrk~590 in detail, and then to investigate the possible mechanism for the rapid change in the dust innermost radius.

In this study, we analyse archival multi-epoch optical photometry data for Mrk~590 obtained in $1998-2007$ by the Sloan Digital Sky Survey (SDSS).
Using these multi-epoch data, we constrain the period when the optical luminosity of Mrk~590 had begun to decrease, which was suggested by \cite{den14} to have occurred sometime between 1996 and 2003.
We show that Mrk~590 experienced a sudden, large luminosity decrease during a brief period from 2000 to 2001.
Then, from the time difference between the epoch of the sudden luminosity decrease and the epoch of the dust lag measurement by MAGNUM, we constrain the time scale within which the innermost radius of the dust torus of Mrk~590 had been adjusted to the decreased dust sublimation radius in the faint state (Section~\ref{sec:expected_dust}).
We conclude that the variations in the dust innermost radius of Mrk~590 occurred so rapidly that they cannot be explained by radial inflow of the dust clouds (Section~\ref{sec:reform}).
Instead, we suggest that the observed short time scale of the variations in the dust innermost radius requires new dust formation  either in the radiatively-cooled BLR gas located in between dust sublimation radii at the bright and faint states of Mrk~590 (Section~\ref{sec:dust_in_blr}) or in the cooled disk atmosphere (Section~\ref{sec:dust_in_disk}).

Throughout this paper, we assume 3-Year Wilkinson Microwave Anisotropy Probe cosmology \citep{spe07}; $H_0 = 73$~km~s${}^{-1}$~Mpc${}^{-1}$, $\Omega_{m} = 0.27$, and $\Omega_{\Lambda}=0.73$.
The zero-point flux of the AB magnitude system is 3,631~Jy, and the SDSS magnitude system is assumed to be identical to the AB magnitude system \citep[e.g.,][]{suz18}.
Unless otherwise stated, in the following the flux values listed are uncorrected for Galactic extinction, whereas the luminosities are corrected for Galactic extinction using the extinction coefficients of \cite{sch11}, following the custom of the previous works \citep{pet98b,kos14}.

\section{Data}
\label{sec:2}

In this section, we first describe the basic properties of Mrk~590 (Section~\ref{sec:mrk590}) and historical measurements of the BLR and dust reverberation lags of Mrk~590 and other Seyfert galaxies to illustrate that Mrk~590 is an outlier for the global relationship between  $R_{\text{dust}}$ and $R_{\text{H}\beta}$ (Section~\ref{blr_dust_radii}).
We then describe details of the historical optical and NIR photometry data used in this paper, and evaluate aperture fluxes and host galaxy flux contributions within the aperture to determine flux variation of the AGN component during the observation period [MAGNUM data in Section~\ref{magnum_obs}, data from \cite{pet98b} in Section~\ref{peterson_obs}, SDSS data in Section~\ref{sec:sdss_photometry}, and 2MASS and DENIS data in \ref{sec:nirdata_2mass_denis}].

\subsection{Mrk~590}
\label{sec:mrk590}

Mrk~590 (R.A = 02:14:33.561, Decl. = $-$00:46:00.18), a ``changing-look'' AGN at $z=0.02639$, was a typical Seyfert 1 AGN before 2006, but has since been reclassified as a Seyfert $1.9-2$ owing to the disappearance of the broad Balmer emission lines in the optical spectra after 2013 \citep{pet98b,lan08,ben09,den14}.
The luminosity distance and angular scale corrected for the Virgo infall $+$ Great Attractor $+$ Shapley supercluster local flow are $d_L = 107$~Mpc and 0.495~kpc~arcsec${}^{-1}$, respectively (taken from the NASA/IPAC Extragalactic Database, NED).

The H$\beta$ BLR reverberation mapping measurement provides an estimate of $M_{\rm BH} =3.71^{+0.57}_{-0.58} \times 10^{7} M_{\odot}$ \citep{gri13,ben15}.
The Galactic extinction coefficients toward the direction of Mrk~590 in units of magnitude are as follows:
$A_V = 0.101$, 
$A_g = 0.121$,
$A_r = 0.084$, and
$A_K = 0.011$~mag \citep[taken from NED, originally from][]{fit99,sch11}.

\subsection{BLR and dust reverberation radii: Mrk~590 as an outlier}
\label{blr_dust_radii}

\begin{table*}
\centering
\footnotesize
%\scriptsize
\caption{Rest-frame H$\beta$ and dust reverberation lag measurements for 15 dust reverberation-mapped Seyfert galaxies.}
\label{tbl-4}
\begin{tabular}{l c c c c c}
\hline
(1)   & (2)                &     (3)            & (4)     & (5) & (6) \\
Name & $\tau_{\rm rest, H\beta~BLR}$ & $\log L_{5100, \text{AGN}}$ ($H\beta$ lag) & Reference & $\tau_{\rm rest, dust}$ & $\log L_{5100, \text{AGN}}$ (dust lag)\\
 & (days) & (erg~s${}^{-1}$) &  & (days) & (erg~s${}^{-1}$)\\
\hline
Mrk 335         & ${15.7  }_{-4.0   }^{+3.4}$ & ${43.71}\pm0.01$ & (1,5)   & ${138.6}\pm 16.3$&         $43.63\pm0.01$\\
Akn 120         & ${39.7  }_{-5.5   }^{+3.9}$ & ${43.69}\pm0.06$ & (1,5)   & ${134.5}\pm 17.0$&         $44.25\pm0.01$\\
MCG+08-11-011  & ${15.72 }_{-0.52  }^{+0.5}$ & ${43.26}       $ & (2)     & ${ 93.9}\pm 13.5$&         $43.43\pm0.02$\\
Mrk 79          & ${15.2  }_{-5.1   }^{+3.4}$ & ${43.62}\pm0.01$ & (1,5)   & ${ 72.4}\pm  3.5$&         $43.36\pm0.02$\\
Mrk 110         & ${25.5  }_{-5.6   }^{+4.2}$ & ${43.61}\pm0.01$ & (1,5)   & ${ 87.2}\pm  5.9$&         $43.63\pm0.02$\\
NGC 3227        & ${3.75  }_{-0.82  }^{+0.76}$& ${42.11}\pm0.05$ & (3,5)   & ${ 14.4}\pm  0.6$&         $42.20\pm0.02$\\
NGC 3516        & ${11.68 }_{-1.53  }^{+1.02}$& ${42.82}\pm0.09$ & (3,5)   & ${ 52.5}\pm  9.4$&         $42.63\pm0.04$\\
NGC 4051        & ${1.87  }_{-0.5   }^{+0.54}$& ${41.72}\pm0.06$ & (3,5)   & ${ 14.7}\pm  0.5$&         $41.68\pm0.04$\\
NGC 4151        & ${6.6   }_{-0.8   }^{+1.1}$ & ${42.25}\pm0.09$ & (1,5)   & ${ 49.5}\pm  0.7$&         $42.63\pm0.04$\\
NGC 4593        & ${3.7   }_{-0.8   }^{+0.8}$ & ${42.96}\pm0.03$ & (1,5)   & ${ 43.1}\pm  1.8$&         $42.55\pm0.02$\\
NGC 5548        & ${12.4  }_{-3.85  }^{+2.74}$& ${43.02}\pm0.06$ & (3,5)   & ${ 54.3}\pm  0.7$&         $42.89\pm0.01$\\
Mrk 817         & ${14.04 }_{-3.47  }^{+3.41}$& ${43.80}\pm0.01$ & (3,5)   & ${ 87.1}\pm  8.0$&         $43.73\pm0.01$\\
Mrk 509         & ${79.6  }_{-5.4   }^{+6.1}$ & ${44.13}\pm0.01$ & (1,5)   & ${144.2}\pm  8.9$&         $44.22\pm0.01$\\
NGC 7469        & ${10.8  }_{-1.3   }^{+3.4}$ & ${43.43}\pm0.03$ & (4)     & ${ 47.5}\pm  1.4$&         $43.28\pm0.02$\\
Mrk 590         & ${25.6  }_{-2.3   }^{+2.0}$ & ${43.31}\pm0.05$ & (1,5)   & ${ 36.2}\pm  2.6$&         $42.85\pm0.03$\\\hline
Mrk 590 (epoch 1, MJD=48090-48323)  & $20.7^{+3.5}_{-2.7}$ &  $43.49 \pm 0.04$ & (1,5)   &                  &              \\
Mrk 590 (epoch 2, MJD=48848-49048)  & $14.0^{+8.5}_{-8.8}$ &  $43.03 \pm 0.11$ & (1,5)   &                  &              \\
Mrk 590 (epoch 3, MJD=49183-49338)  & $29.2^{+4.9}_{-5.0}$ &  $43.28 \pm 0.06$ & (1,5)   &                  &              \\
Mrk 590 (epoch 4, MJD=49958-50122)  & $28.8^{+3.6}_{-4.6}$ &  $43.55 \pm 0.03$ & (1,5)   &                  &              \\
 \hline
\end{tabular}
\\
\begin{flushleft}
Column (2) is the rest-frame BLR H$\beta$ lags.
Column (3) is the starlight-corrected AGN luminosities at the rest-frame 5100~\AA~at the epochs of the H$\beta$ BLR lag measurements.
The luminosities are recalculated using luminosity distances on the assumption of $\Lambda$CDM cosmology of $H_0=73$~km~s${}^{-1}$~Mpc${}^{-1}$, $\Omega_{m}=0.27$, and $\Omega_{\Lambda}=0.73$ (the Virgo infall + Great Attractor + Shapley supercluster local flow) and the extinction coefficients of \cite{sch11} obtained from NED.
The uncertainty in the luminosity does not include the uncertainty in the distance.
For objects with multiple lag measurements, weighted averages of H$\beta$ lags and AGN luminosities are listed.
Column (4) shows the references of the BLR H$\beta$ reverberation measurements.
Column (5) shows the weighted average cross-correlation centroid dust lags (Table~7, Column 5 of \cite{kos14}), corrected for a redshift dilation of $1+z$ to convert them to rest-frame values.
The cross-correlation was calculated by assuming $\alpha_{\nu} = 0$ for the AGN continuum to subtract the accretion-disk component in the $K$-band flux.
Column (6) is the AGN luminosities at the epochs of the dust lag measurements (Table~7, Column 4 of \cite{kos14}), on the same assumption regarding the cosmology and extinction coefficients of Column (3).\\
References --- (1) \cite{ben09}, (2) \cite{fau17b}, (3) \cite{den10}, (4) \cite{pet14}, (5) \cite{ben13}.
\end{flushleft}
\end{table*}

Among the 17 AGNs for which dust reverberation lags were measured by the MAGNUM project \citep{kos14}, 15 have H$\beta$ BLR lag measurements in the literature, as listed in Table~\ref{tbl-4}.
Most of the H$\beta$ reverberation lag measurements in Table~\ref{tbl-4} are taken from \cite{ben09,ben13}, which provide the most up-to-date compilation of these measurements \citep{pet98b,pet04,ben06b,den06,den10}.
For objects with multiple lag measurements, weighted averages of H$\beta$ lags taken from \cite{ben09} along with AGN luminosities calculated from the AGN fluxes taken from \cite{ben13} are listed in Table~\ref{tbl-4}.
For NGC~7469 and MCG+08-11-011, recent H$\beta$ BLR reverberation mapping results are taken from \cite{pet14} and \cite{fau17b}, respectively.
In Table~\ref{tbl-4}, the individual H$\beta$ lag and rest-frame 5100~\AA\ AGN luminosity of Mrk~590 at each of the four epochs of the reverberation mapping observations are also shown.
The four epochs of the H$\beta$ lags and the AGN rest-frame 5100~\AA\ luminosities are based on light curve data collected by \cite{pet98b}, whose MJD ranges are $48090-48323$ (epoch 1), $48848-49048$ (epoch 2), $49183-49338$ (epoch 3), and $49958-50122$ (epoch 4).

\begin{figure}
\center{
\includegraphics[clip, width=3.2in]{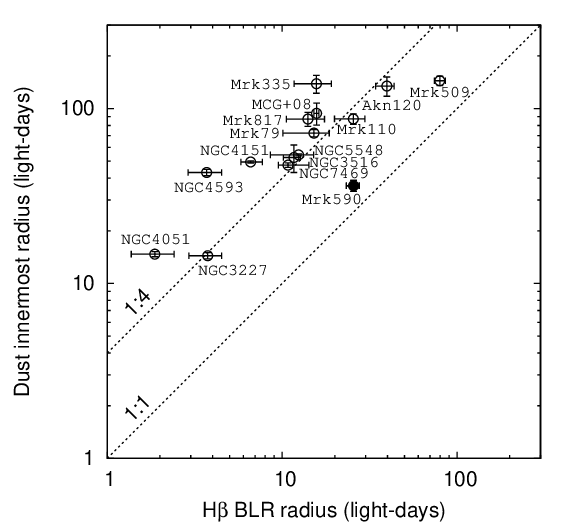}
\includegraphics[clip, width=3.2in]{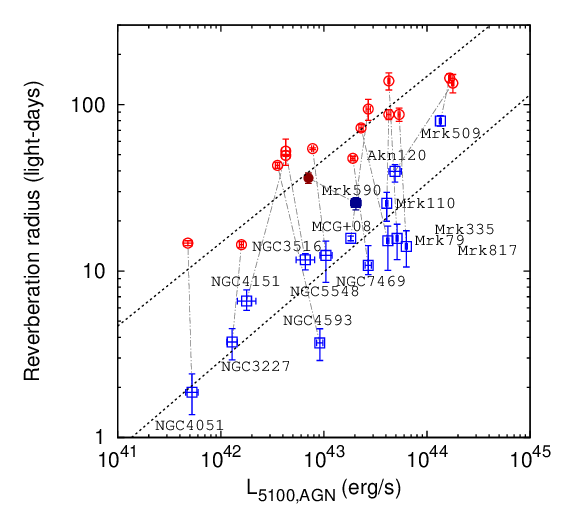}
}
 \caption{Upper panel: the relationship between the broad line region (BLR) radius and the dust innermost radius for the 15 active galactic nuclei (AGNs) for which both the H$\beta$ and dust reverberation lags have been measured (Table~\ref{tbl-4}).
 Mrk~590 is shown by a filled circle.
 Dotted lines denote $R_{{\rm dust}} = R_{{\rm H}\beta}$ and $R_{{\rm dust}} = 4 R_{{\rm H}\beta}$.
 Lower panel: the lag$-$luminosity plots for the dust innermost radius (open circles) and H$\beta$ BLR radius (open squares). 
 The best-fit dust and H$\beta$ BLR lag$-$luminosity relationship obtained by Bentz et al. (2013) and Koshida et al. (2014), respectively, are denoted by dotted lines. $R_{{\rm dust}}$ and $R_{{\rm H}\beta}$ of the same object are connected by dot-dash lines. 
 Mrk~590 is highlighted by filled symbols.
 }
 \label{fig:dust_blr}
\end{figure}

The upper panel of Fig.~\ref{fig:dust_blr} shows the relationship between the BLR radius and the dust innermost radius of 15 Seyfert galaxies determined from reverberation mapping observations, as summarised in Table~\ref{tbl-4}.
The dust radii generally have a larger radius compared to the H$\beta$ BLR radius, by a factor of $\sim 4$, as already pointed out by several authors \citep{sug06,kos14,du15,ben16,bas18}; this is most likely a natural consequence of the sublimation of dust grains in dense gas ($n\sim 10^{10}$~cm${}^{-3}$), with a typical grain size of $\sim 0.01-0.1$~$\mu$m \citep[][]{yos14,bas18}.

In the upper panel of Fig.~\ref{fig:dust_blr}, we can also identify that Mrk~590 is a clear outlier from the global BLR-dust radius relationship.\footnote{Mrk~509 (not to be confused with Mrk~590) is also an outlier in the upper panel of Fig.~\ref{fig:dust_blr}, and both the dust and BLR radii slightly deviate from the global radius$-$luminosity relationships, such that they are closer to each other, as presented in the lower panel of Fig.~\ref{fig:dust_blr}. The expected dust reverberation lag of Mrk~509 ($\sim 190-200$~days; bottom panel of Fig.~\ref{fig:dust_blr}) roughly corresponds to a seasonal gap; thus, the dust lag measurement probably suffers from aliasing effects, due to the seasonal gap in the light curves \citep[e.g.,][]{zu11}. The possible reasons for the unexpectedly large BLR radius compared to the global radius-luminosity relation have yet to be resolved.}
The H$\beta$ radius of Mrk~590 is only slightly larger than the dust radius (by a factor of $\sim 1.4$), as pointed out by \cite{gan15} and \cite{du15}.

In the lower panel of Fig.~\ref{fig:dust_blr}, the BLR and dust reverberation radii as a function of rest-frame 5100~\AA\ AGN luminosity at the epoch of the reverberation lag measurement are plotted for the 15 AGNs.
Although Mrk 590 is the clear outlier in the BLR-dust radius relationship (upper panel of Fig.~\ref{fig:dust_blr}), its dust and BLR reverberation radii follow their respective global correlations with luminosity.
In fact, the AGN luminosity of Mrk~590 at the epoch of the dust reverberation measurement is much smaller than that of the BLR reverberation measurement.
This suggests that the unexpectedly small dust innermost radius of Mrk~590, compared to the BLR radius, can be interpreted as a result of the reduced dust innermost radius, according to the significant UV-optical luminosity drop that occurred after the epoch of the BLR reverberation lag measurement.
The aim of this work is to examine the variation of the innermost radius of the
dust distribution of Mrk 590 as a function of the AGN luminosity variation in more detail.

\subsection{$V$- and $K$-band light curves of Mrk~590 during the MAGNUM dust reverberation mapping observations from 2003$-$2007}
\label{magnum_obs}

The MAGNUM project conducted long-term monitoring observations in optical and NIR wavelengths for a number of type 1 AGNs, to determine the dust reverberation lags in the AGNs \citep{sug06,kos14}.
MAGNUM observations were carried out using a dedicated 2-m MAGNUM telescope and an optical-NIR simultaneous imaging camera \citep{kob98,kob98b}.

In this study, we use the MAGNUM $V$- and $K$-band light curves of Mrk~590 at MJD $=$ $52642-54320$ (from 2003 January 3 to 2007 August 8) given in Table~4 of \cite{kos14}.
Fig.~\ref{fig:mrk590_MAGNUM_lc} shows the Galactic extinction uncorrected $V$- and $K$-band light curves of Mrk~590.
\footnote{Although \cite{kos14} mentioned that the Galactic extinction is uncorrected for the light curve of Mrk 590 presented in their Table~4, we compare $V$-band data with those presented in \cite{sak10} and find that the Galactic extinction is actually corrected for them based on the extinction coefficients of \cite{sch98}. Therefore, we apply the Galactic extinction to the $V$- and $K$-band fluxes of Mrk~590 from \cite{kos14}'s Table~4 to obtain the Galactic extinction uncorrected light curves.}
During the observations of MAGNUM, Mrk~590 showed correlated $V$- and $K$-band flux variability \citep[see][for details]{kos14}.
As noted by \cite{den14}, although the BELs had already become very weak at the epochs of the MAGNUM observations, the optical continuum variability behaviour of Mrk~590 was normal for a Seyfert 1 galaxy of this luminosity and black hole mass, suggesting that there was no significant change in the accretion disk state, even in the faint phase.

The MAGNUM AGN photometry presented in \cite{kos14} is obtained using a circular aperture with a diameter of $\phi = 8''.3$.
Sky flux is estimated from the flux within a $\phi = 11''.1-13''.9$ annulus aperture centred on the AGN \citep{sak10,kos14}.
In the case of Mrk~590, the sky annulus corresponds to galactocentric distances of $5.5-6.9$~kpc; thus, the sky flux is over-subtracted, due to the flux contribution from the host galaxy stellar disk component.
The Galactic extinction uncorrected host galaxy flux estimates given in \cite{kos14}, $f_{V, \text{host}} = 4.24 \pm 0.06$~mJy and $f_{K, \text{host}} = 22.18 \pm 0.13$~mJy \citep[originally derived by][]{sak10}, are evaluated from point spread function (PSF)-subtracted coadded MAGNUM images using the same circular aperture and sky annulus aperture.
As described in Section~\ref{sec:host_estimate_acs}, we have re-evaluated the $V$-band host galaxy flux contribution to the MAGNUM aperture from a high spatial resolution HST/ACS F550M image as $f_{V, \text{host}} = 4.396 \pm 0.015$~mJy, on the assumption of a colour correction based on the bulge template spectrum presented in the study by \cite{kin96}; throughout this work, we use this re-evaluated value as the $V$-band host galaxy flux.

The weighted averages of the MAGNUM $V$- and $K$-band light curves of Mrk~590 are $f_{\nu}(V) = 4.669 \pm 0.002$~mJy and $f_{\nu}(K) = 24.188 \pm 0.019$~mJy, respectively (the Galactic extinction is uncorrected).
By subtracting the host galaxy flux contributions of $f_{V, \text{host}} = 4.396  \pm 0.015$~mJy and  $f_{K, \text{host}} = 22.18 \pm 0.13$~mJy, the average AGN $V$-band continuum flux is $f_{\nu} (V, {\rm AGN}) = 0.27 \pm 0.02$~mJy and $f_{\nu} (K, {\rm AGN}) = 2.01 \pm 0.13$~mJy, respectively.

\subsection{Optical continuum light curves during the H$\beta$ BLR reverberation mapping observations from 1990$-$1996}
\label{peterson_obs}

The H$\beta$ BLR reverberation mapping observations for Mrk~590 were carried out by \cite{pet98b} in 1990$-$1996.
The spectroscopic continuum light curves at $\lambda_{\text{rest}} = 5100$~\AA\ ($\lambda_\text{obs} = 5,240-5,260$~\AA) presented in \cite{pet98b} are obtained with a $5''$-width slit and extracted through a $7''.6$ aperture.
We directly use these spectroscopic continuum light curve data of Mrk~590 to examine the flux variation of Mrk~590 at the epoch of the BLR reverberation measurement.
The light curve data were downloaded from the Ohio State AGN Spectroscopic Monitoring Project website.\footnote{\href{http://www.astronomy.ohio-state.edu/~peterson/AGN/}{http://www.astronomy.ohio-state.edu/~peterson/AGN/}}

Fig.~\ref{fig:mrk590_MAGNUM_lc} shows the starlight-uncorrected spectroscopic continuum light curve at $\lambda_{\text{rest}} = 5100$~\AA\ ($\lambda_\text{obs} = 5,240-5,260$~\AA) during the H$\beta$ BLR reverberation mapping observations.
As can be seen from this figure, the flux variability amplitude was larger during the H$\beta$ RM observations of \cite{pet98b} in the 1990s, compared to that during the MAGNUM dust RM observations in the 2000s.
\cite{ben13} estimate the host galaxy flux contribution to the spectroscopic aperture of \cite{pet98b} as $f_{\lambda, 5100(1+z)~\text{\AA}, \text{host}} = 3.965 (\pm 0.198) \times 10^{-15}$~erg~s${}^{-1}$~cm${}^{-2}$~\AA${}^{-1}$, and derive the mean AGN fluxes at $\lambda_{\text{rest}} = 5100$~\AA~as 
$f_{\lambda, 5100(1+z)~\text{\AA}, \text{AGN}} = 
3.93\pm0.33$, 
$1.37\pm0.31$, 
$2.40\pm0.31$, and 
$4.46\pm0.34$ 
$\times$ $10^{-15}$~erg~s${}^{-1}$~cm${}^{-2}$~\AA${}^{-1}$ 
at MJD = $48090-48323$, $48848-49048$, $49183-49338$, and $49958-50122$, respectively.
Assuming the power-law index of $\alpha_{\nu} = 0$ for the AGN continuum based on the spectral energy distribution (SED) of the variable component \citep{sak10}, $f_{\lambda, 5100(1+z)~\text{\AA}, \text{AGN}}$ can be converted into the mean AGN fluxes at $V$-band as
$f_{\nu} (V, {\rm AGN}) = 
3.61\pm0.30, 
1.26\pm0.29, 
2.21\pm0.29, 
4.10\pm0.31$~mJy 
at MJD = $48090-48323$, $48848-49048$, $49183-49338$, and $49958-50122$, respectively.
Compared with the weighted average AGN $V$-band flux $f_{\nu} (V, {\rm AGN}) = 0.27 \pm 0.02$~mJy during the MAGNUM observations (Section~\ref{magnum_obs}), we can say that the AGN flux of Mrk~590 was about 10 times brighter in the 1990s compared to the epochs of the MAGNUM observations in the 2000s.

\subsection{SDSS Stripe~82 multi-epoch optical photometry in 1998$-$2007}
\label{sec:sdss_photometry}

\begin{figure}
\center{
\includegraphics[clip, width=3.7in]{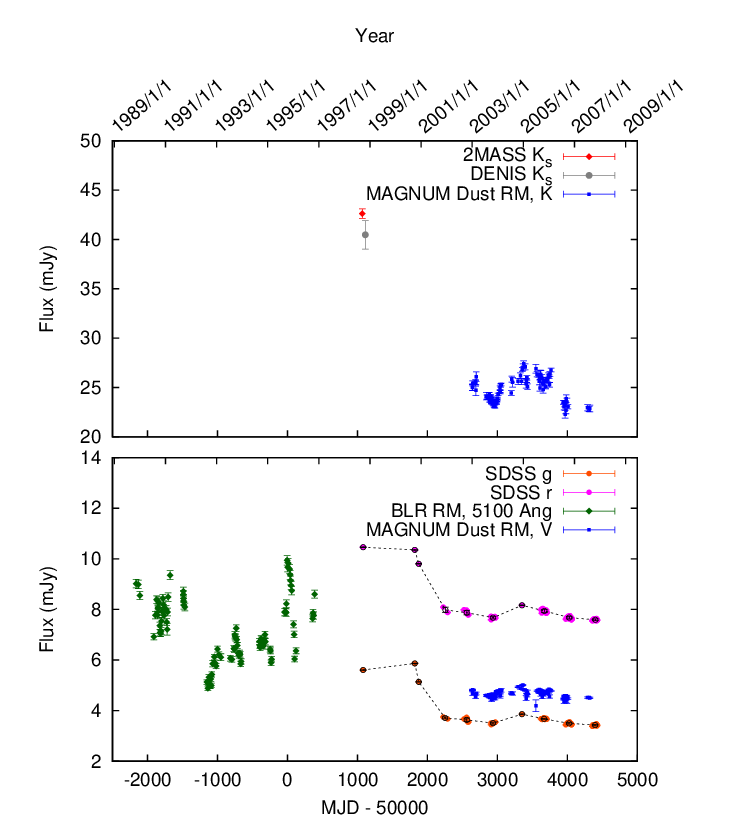}
}
 \caption{
 Top: the Two Micron All-Sky Survey (2MASS), Deep Near Infrared Survey of the Southern Sky (DENIS), and Multicolor Active Galactic Nuclei Monitoring (MAGNUM) $K$-band light curve of Mrk~590.
 Bottom: the Sloan Digital Sky Survey (SDSS) Stripe~82 $g$- and $r$-band light curves (filled circles: unbinned; open circles: binned) and MAGNUM $V$-band light curves of Mrk~590. 
 The 5100~\AA\ fluxes from Peterson et al. (1998)'s H$\beta$ BLR reverberation mapping (RM) data are shown for comparison.
 The Galactic extinction is uncorrected, and the host galaxy flux contributions are not subtracted.
 Mrk~590 was in bright phase before 2000 and then suddenly transitioned into a faint phase during 2000$-$2001.
 }
 \label{fig:mrk590_MAGNUM_lc}
\end{figure}

Mrk~590 is located inside of the SDSS Stripe~82 region, which was continuously observed with the SDSS telescope in $u$-, $g$-, $r$-, $i$-, and $z$-band filters over the course of the SDSS-I/-II Stripe 82 programs from 1998 to 2007 \citep[][]{fuk96,gun98,yor00,bra08,fri08,sak18}\footnote{\href{https://classic.sdss.org/dr7/coverage/sndr7.html}{https://classic.sdss.org/dr7/coverage/sndr7.html}}.

In general, the SDSS Stripe~82 images obtained after September~2005 were not taken under photometric conditions, with less optimal seeing, moon-phase, and photometric conditions compared to the main SDSS Legacy Survey data obtained prior to September~2005 \citep[see][, for details regarding the photometric calibration for the Stripe~82 images]{ann14}.
Since Mrk~590 has an extended morphology in the optical images (Section~\ref{sec:host_estimate}), the PSF photometry (and other model photometry) magnitudes of Mrk~590 recorded in SDSS photometry catalogues produced by SDSS pipelines cannot be used directly to examine the AGN flux variation due to variable aperture losses under the unstable seeing conditions.
To manage and improve photometry under less optimal conditions, in this work we use the $g$- and $r$-band SDSS Stripe 82 images and apply aperture photometry with a fixed large aperture size to them instead of using the processed photometric data in the catalogue.

The SDSS corrected frame images ({\tt fpC}), including the sky region of Mrk~590 and associated field information tables ({\tt tsField}), were downloaded from the SDSS Data Release (DR) 7 Data Archive Server.
Each of the {\tt fpC} images has an image size of $1,489 \times 2,048$ pixels, with a pixel scale of $0''.396$~pixel${}^{-1}$.
Mrk~590 was imaged in 79 drift scan runs.
For each run, two images with adjacent field numbers overlap along the scan direction by 128 pixels. Mrk~590 is located in the overlapping region for runs 2709, 4203, 5681, and 5776; for these runs, we only analysed one of two images with field numbers 92, 630, 80, and 629, respectively.
We visually checked the images and excluded images obtained as runs 5776, 5871, and 6349 from the analysis below, due to significant residual background patterns.
Images obtained for run 5637 were excluded from the analysis, as the associated {\tt tsField} files were \href{http://das.sdss.org/raw/5637/40/calibChunks/2/knownMissing.txt}{missing}.
In addition, the images taken under seeing full width half maximums (FWHMs) of $> 2$~arcsec (judged using {\tt psf\_width} recorded in {\tt tsField}) were excluded from the analysis.
For the $g$- and $r$-band analysis, 58 and 64 drift scans are retained, respectively.

First, global sky background subtraction and $\phi 8''.3$ aperture photometry for field stars are performed using {\tt SExtractor} \citep{ber96}.
Here, we adopt the aperture diameter size of $\phi 8''.3$ to match the standard MAGNUM photometry (Section~\ref{magnum_obs}).
The MAGNUM photometry uses $\phi 11''.1-13''.9$ annulus aperture to estimate the sky flux to be subtracted, whereas our SExtractor analysis of the SDSS images estimates global sky background using $128 \times 128$ pixels background meshes.
The background sigma map is calculated by the {\tt SExtractor}, and object Poisson noise contribution is calculated using charge-coupled device (CCD) gain values of 3.855 and 4.600 for the $g$- and $r$-bands, respectively.
The zero-point magnitude of each image is determined by comparing the measured instrumental aperture magnitudes and the SDSS PSF magnitudes of field objects catalogued in the SDSS DR7 Catalog Archive Server.
Then, aperture fluxes of Mrk~590 and their statistical errors are measured using {\tt Photutils} \citep{pho17} adopting a $\phi8''.3$ circular aperture centred on the sky coordinate of the nucleus of Mrk~590.

The SDSS Stripe~82 observations after 2001 generally obtained multiple images within each year, during which the AGN variability of Mrk~590 was small (see below). 
A few outlying photometry data points are identified and removed from the light curves by applying $3\sigma$ clipping for photometry data obtained within a year (only for the years with over three data points), where $\sigma$ is defined as the median absolute deviation of the data points.
After rejection of the outliers (56 and 64 data points are retained for the $g$- and $r$-bands, respectively), photometry data points obtained within a year are binned by taking a weighted mean, and the median absolute deviation of the data points are assigned as the photometric error for the binned data.

Fig.~\ref{fig:mrk590_MAGNUM_lc} presents the $\phi 8''.3$ aperture SDSS $g$- and $r$-band light curves.
SDSS observations reveal that Mrk~590 experienced a sudden flux decrease in 2000-2001 (MJD = $51819-52288$).
Hereafter, we refer to the period after the significant flux decrease as the ``faint phase'' of Mrk~590, and the period before 2000 as the ``bright phase''.
Previously, a significant AGN optical luminosity drop in Mrk~590 from 1996 to 2013 was identified using discrete spectroscopic data obtained in 1996, 2003, 2006, and 2013 \citep{den14}; the continuous SDSS imaging data analysed here pinpoint the exact epoch of the beginning of the significant luminosity drop to be between 2000 and 2001.

Despite the significant luminosity drop in 2000$-$2001, the SDSS spectrum of Mrk~590 obtained on 2003 January 10 (MJD = 52649), i.e. in the faint phase, still showed broad H$\beta$ and H$\alpha$ emission line features \citep{den14}.
Moreover, the broad hydrogen Balmer, Paschen, and Bracket emission lines were clearly detected in the optical and NIR spectra obtained by \cite{lan08,lan11} during $2006-2007$.
This indicates that the changing-look of Mrk~590 had not occurred immediately after the significant luminosity drop between 2000 and 2001; instead, the subsequent luminosity decrease in the 2000s to 2010s eventually led Mrk~590 to become extremely faint, leading to the type~2-like optical spectral state revealed by optical spectroscopy data obtained in 2013 \citep[see Figure~4 of][]{den14}.

In the binned light curves shown in Fig.~\ref{fig:mrk590_MAGNUM_lc}, the lowest observed fluxes during the SDSS observations are $3.428 \pm 0.032$ and $7.590 \pm 0.034$~mJy in the $g$ and $r$ bands, respectively, in 2007.
Here, the quoted uncertainties are the standard deviations of the observed data points within each year; thus, they include the host galaxy flux uncertainties due to the variable seeing effects on aperture photometry.
The host galaxy flux contributions to the $\phi 8''.3$ aperture must be less than these observed fluxes, and they are consistent with the estimates of the host galaxy flux contributions of $3.373 \pm 0.029$~mJy and $7.538 \pm 0.047$~mJy at the $g$ and $r$ bands, respectively, from the GALFIT modelling, as described in Section~\ref{sec:host_estimate_sdss}.
The AGN fluxes at this epoch are therefore estimated as $0.055 \pm 0.043$~mJy and $0.052 \pm 0.058$~mJy at the $g$ and $r$ bands, respectively, indicating that the AGN flux contributions were very low at this epoch (see Section~\ref{sec:expected_dust} for further details).

Subtracting the host galaxy flux contributions from the $g$- and $r$-band light curves, we can see that the AGN optical emission of Mrk~590 decreased by a factor of $\gtrsim 10$ during the SDSS observations (see Section~\ref{sec:expected_dust} for further details).

\subsection{2MASS and DENIS near-infrared photometry in 1998}
\label{sec:nirdata_2mass_denis}

The Two Micron All-Sky Survey \citep[2MASS;][]{coh03,skr06} database provides $J$- (1.24~$\mu$m), $H$- (1.66~$\mu$m), and $K_s$- (2.16~$\mu$m) band all-sky images obtained with three-band simultaneous NIR imaging cameras mounted on dedicated 1.3-m telescopes.
Mrk~590 was imaged by 2MASS on 1998 September 13 (MJD = $51069.3$) by the 2MASS telescope at the Cerro Tololo Inter-American Observatory (CTIO), near La Serena, Chile, when Mrk~590 was in the bright phase according to SDSS observations.
In the same year, the Deep Near Infrared Survey of the Southern Sky \citep[DENIS;][]{fou00,den05} also observed Mrk~590 at Gunn $i$ (0.79~$\mu$m), $J$ (1.23~$\mu$m), and $K_{s}$ (2.15~$\mu$m) bands (1998 Oct. 26, MJD = 51112.2), using a three-band camera mounted on the 1-m European Southern Observatory (ESO) telescope at La Silla (Chile).

We retrieved $J$-, $H$-, and $K_s$-band Atlas images ($512 \times 1,024$ pixels, $1''$~pixel${}^{-1}$) at the position including Mrk~590 from 2MASS All-Sky Release Data Products at the NASA/IPAC (Caltech, CA, USA) Infrared Science Archive \footnote{\href{http://irsa.ipac.caltech.edu/applications/2MASS/IM/interactive.html}{http://irsa.ipac.caltech.edu/applications/2MASS/IM/interactive.html}} and performed aperture photometry on the Atlas images using {\tt SExtractor} \citep{ber96} and {\tt Photutils} \citep{pho17}.
The same aperture size and sky annulus with MAGNUM photometry, i.e. $\phi 8''.3$ diameter aperture and $\phi 11''.1-13''.9$ sky annulus aperture (Section~\ref{magnum_obs}), are used.
The zero-point magnitude of each image is evaluated by comparing the instrumental aperture magnitudes with the magnitudes of field stars catalogued in the 2MASS All-Sky Point Source Catalogue (PSC), and then the magnitudes are converted to flux units using the zero-magnitude fluxes taken from Table~2 of \cite{coh03}.
Following the standard analysis procedures described in 2MASS web pages \footnote{\href{https://ipac.caltech.edu/2mass/releases/allsky/doc/sec6_8a.html}{https://ipac.caltech.edu/2mass/releases/allsky/doc/sec6\_8a.html}}, additional noise due to coaddition, pixel resampling, and background fit is taken into account.
The 2MASS aperture fluxes of Mrk~590 at the $J$-, $H$-, and $K_s$-bands are measured to be $22.31 \pm 0.21$, $32.16 \pm 0.33$, and $42.62 \pm 0.48$~mJy, respectively (Galactic extinction is not corrected).

In the same manner as for the 2MASS data analysis, we retrieved $i$-, $J$-, and $K_s$-band DENIS images at the position including Mrk~590 from the DENIS public server\footnote{\href{http://cdsweb.u-strasbg.fr/denis.html}{http://cdsweb.u-strasbg.fr/denis.html}}, and aperture photometry was performed using {\tt SExtractor} and {\tt Photutils}.
The zero-point magnitude of each image is evaluated by comparing the instrumental aperture magnitudes with field stars' PSF magnitudes catalogued in the Third DENIS data release catalogue \citep{den05}; the magnitudes are then converted to flux units using the zero-magnitude fluxes taken from Table~4 of \cite{fou00}.
The DENIS aperture fluxes of Mrk~590 at the $i$-, $J$-, and $K_s$-bands are $12.11 \pm 0.12$,  $22.31 \pm 0.44$, and $40.47 \pm 1.44$~mJy, respectively (Galactic extinction is not corrected).
The DENIS $K_{s}$-band magnitude is consistent with the 2MASS $K_{s}$-band magnitude, suggesting that Mrk~590 remained bright during 1998.

Using the host galaxy flux contribution to the $K$-band estimated by \cite{kos14}, $f_{\nu}(K, \rm host) = 22.18 \pm 0.13$~mJy, the AGN component at the 2MASS $K_{s}$-band can be estimated as $f_{\nu}(K_{s}, \rm AGN) = 20.44 \pm 0.50$~mJy.
As described in Section~\ref{magnum_obs}, the AGN $K$-band flux at the epoch of the MAGNUM observation is $f_{\nu} (K, {\rm AGN}) = 2.32 \pm 0.13$~mJy, thus the $K_s$-band AGN emission at the epoch of the MAGNUM observation is about 10 times fainter than that at the epoch of the 2MASS observation (see Fig.~\ref{fig:mrk590_MAGNUM_lc}).
This can naturally be interpreted in terms of a hot dust thermal emission decrease according to significant changes in the disk UV-optical emission that occurred in between the two observations, as traced by SDSS Stripe~82 observations.

\section{Analysis}

In this section, we first revisit the MAGNUM $V$- and $K$-band light curves to put stringent constraints on when the dust reverberation radius was set to $\simeq 30-40$~light-days (lt-days) (Table~\ref{tbl-4}), as inferred by \cite{kos14}.
By using two different lag estimation methods, cross-correlation function (CCF) analysis (Section~\ref{sec:ccf_analysis}) and JAVELIN analysis (Section~\ref{sec:javelin_analysis}), we show that the dust reverberation signal of Mrk~590 was already robustly detected to be $\simeq 30$~light-days in the first year of MAGNUM observations in $2003-2004$ (MJD$ = 52843 - 53055$).
Then, we compare the observed dust reverberation radius with model predictions based on the long-term AGN optical light curves of Mrk~590 and constrain the reformation time scale of the innermost dust distribution after the rapid AGN luminosity decline in $2000-2001$ (Section~\ref{sec:expected_dust}).

\subsection{Dust reverberation lags from the MAGNUM $V$- and $K$-band light curves revisited}
\label{dust_lag_estimate}

\subsubsection{Cross-correlation analysis}
\label{sec:ccf_analysis}

\begin{figure}
\center{
\includegraphics[clip, width=3.5in]{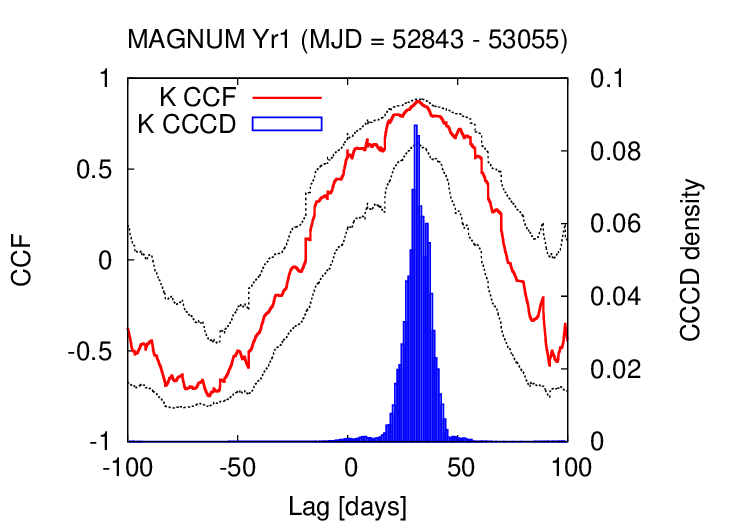}
\par \vspace{0.5cm}
\includegraphics[clip, width=3.5in]{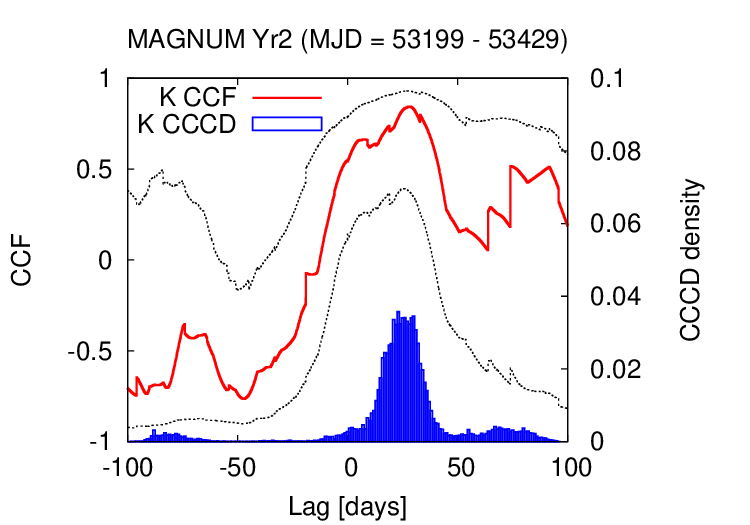}
\par \vspace{0.5cm}
\includegraphics[clip, width=3.5in]{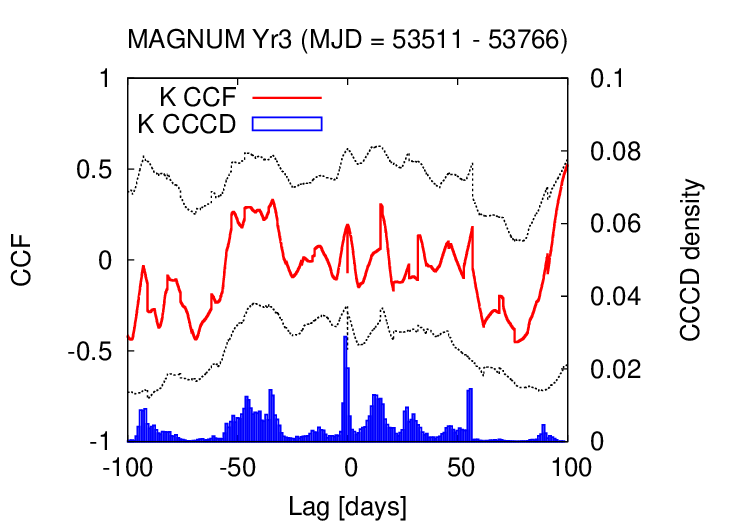}
}
 \caption{Cross-correlation functions (CCFs) and cross-correlation centroid distributions (CCCDs) of the MAGNUM $V$- and $K$-band light curves of Mrk~590 at MAGNUM Yr1 (upper), Yr2 (middle), and Yr3 (bottom) as a function of the observed-frame time lag. Dotted lines denote the 90\% confidence intervals for the CCF. Positive lags indicate that the $K$-band light curve lags behind the $V$-band light curve. The CCCDs
are binned into 1-day bins.}
 \label{fig:ccf_cccd}
\end{figure}

\begin{table*}
\centering
\footnotesize
%\scriptsize
\caption{Rest-frame dust reverberation lags based on cross-correlation analysis and JAVELIN analysis.}
\label{tbl-combinelag}
\begin{tabular}{l c c c}
\hline
(1) & (2) & (3) & (4)\\
MJD & $\tau_\text{rest, dust}$ (CCF) & $\alpha_{\nu}$  (JAVELIN) & $\tau_\text{rest, dust}$ (JAVELIN)\\
 (days)  &  (days)  &    &  (days) \\
\hline
                &                       & $\infty$ & $31.4^{+1.3}_{-1.5}$\\
MAGNUM Yr1: 52843-53055     & $31.6^{+5.4}_{-5.4}$  & $0$      & $32.4^{+1.7}_{-1.5}$\\
                &                       & $1/3$    & $32.0^{+1.3}_{-1.4}$\\\hline
                &                       & $\infty$ & $25.8^{+4.1}_{-3.3}$\\
MAGNUM Yr2: 53199-53429     & $25.2^{+11.8}_{-10.8}$& $0$      & $26.6^{+4.1}_{-3.3}$\\
                &                       & $1/3$    & $26.2^{+4.5}_{-3.4}$\\\hline
\end{tabular}
\\
\begin{flushleft}
(1) MJD ranges of the light curves. (2) Rest-frame dust lags based on CCF analysis; the observed-frame lag divided by a factor of $1+z$. The reported uncertainty is $16-84$\% percentiles. (3) Power-law index models of the variable disk spectrum between the $V$ and $K$ bands assumed in the JAVELIN analysis. (4) the rest-frame dust lags from the JAVELIN analysis. The reported values (and their uncertainties) are 50\% ($16-84$\%) of the posterior distribution.
\end{flushleft}
\end{table*}

The dust reverberation lag of Mrk~590 is measured by \cite{kos14} as $\tau_\text{dust} = 37.2^{+2.7}_{-2.7}$~days ($\tau_\text{rest, dust} = 36.2^{+2.6}_{-2.6}$~days) using a part of the full MAGNUM light curves at MJD = $52842.6-53429.2$ \citep[see Table~7 of][]{kos14}.
To obtain a tighter constraint on the epoch when the dust reverberation radius is set to such a small radius of $\simeq 30-40$~light-days (lt-days), here we re-evaluate the dust reverberation lags of Mrk~590 using the MAGNUM light curves by dividing the light curves more finely on a year-by-year basis, with MJD = $52843-53055$ (MAGNUM Yr1), $53199-53429$ (MAGNUM Yr2), and $53511-53766$ (MAGNUM Yr3).

To infer the dust reverberation lags, the CCF between MAGNUM $V$-band and $K$-band light curves at each epoch is evaluated using an interpolation method, in which the unevenly sampled light curves are linearly interpolated to calculate the correlation coefficients at arbitrary lags \citep{gas86,wel99}.
We use {\tt PYCCF V2} \citep[Python Cross Correlation Function for reverberation mapping studies;][]{sun18b} for the cross-correlation analyses.
The forward CCF is calculated by interpolating the $K$-band light curve to the $V$-band light curve sampling, and the backward CCF is obtained by interpolating the $V$-band light curve to the $K$-band light curve sampling. 
The final CCF is calculated as the average of the two CCFs.
CCFs are obtained at 0.1-day intervals within the range of $-100 \leq \tau_\text{dust, obs}~\text{(days)} \leq +100$, where $\tau_\text{dust, obs}$ denotes the observed-frame lag.
An estimate of the reverberation lag is evaluated as the centroid of the CCF.
The CCF centroid calculation is performed only for CCF points within 80\% of the peak value of the CCF ($0.8 \times \text{CCF}_\text{max}$).
The uncertainties in the CCFs and their centroids are estimated using the flux randomisation and random subsample selection method (FR/RSS) of \cite{wel99}.
Up to $30,000$ Monte Carlo FR/RSS realisations for the $V$- and $K$-band light curve produce a cross-correlation centroid distribution (CCCD) of the dust reverberation lag, whose width is used as an estimate of the measurement uncertainty of the dust reverberation lag.

Fig.~\ref{fig:ccf_cccd} shows the CCFs and CCCDs of the $V$- and $K$-band light curves at MAGNUM Yr1 (top panel), Yr2 (middle panel), and Yr3 (bottom panel); the 90\% confidence intervals of the CCFs calculated from the FR/RSS realisations are also shown.
The CCFs at MAGNUM Yr1 and Yr2 show statistically significant CCF peaks, and positive lags are clearly evident.
The median rest-frame $K$-band time lags and their $16-84$\% percentiles calculated from the CCCDs are summarised in Table~\ref{tbl-combinelag}.
On the other hand, the CCF at MAGNUM Yr3 does not show any statistically significant correlation, and we are unable to obtain the dust reverberation lag at this epoch.

As for the variable component, the $V$-band light curve of Mrk~590 is dominated by disk continuum emission, and the $K$-band light curve is dominated by hot dust emission from the dust torus.
However, the $K$-band flux may include a flux contribution from the long wavelength tail of the disk continuum emission \citep{min06,tom06,kis08,kos14,man18}.
Although the minor contribution of the disk continuum emission in the $K$ band is ignored in the CCF analysis here, the obtained dust lags are consistent with $\tau_\text{rest, dust} = 36.2^{+2.6}_{-2.6}$~day reported by \cite{kos14}, in which the disk continuum emission was subtracted by extrapolating the $V$-band light curve assuming a power-law continuum of $\alpha_{\nu} = 0$.
The above analysis suggests that the dust innermost radius had already become as small as $\simeq 40$~lt-days at the first epoch of the MAGNUM observation at MAGNUM Yr1, which is $\sim$ 1,000~days after the significant decrease in the AGN luminosity of Mrk~590.

\subsubsection{JAVELIN analysis}
\label{sec:javelin_analysis}

\begin{figure*}
\center{
\includegraphics[clip, width=3.2in]{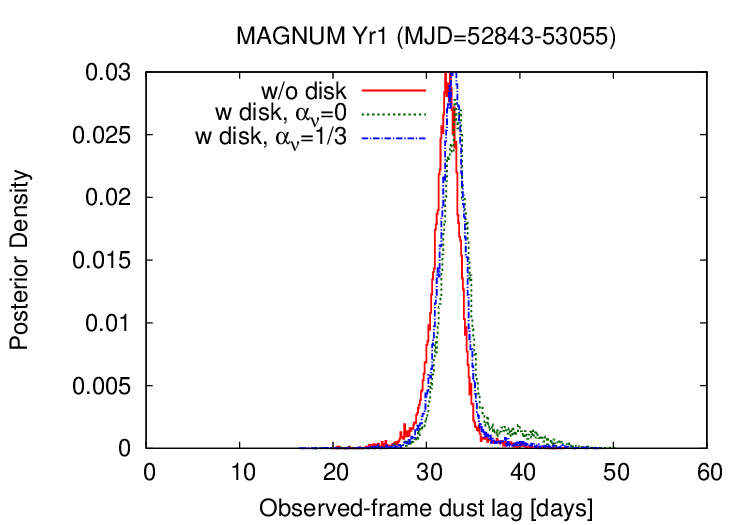}
\includegraphics[clip, width=3.4in]{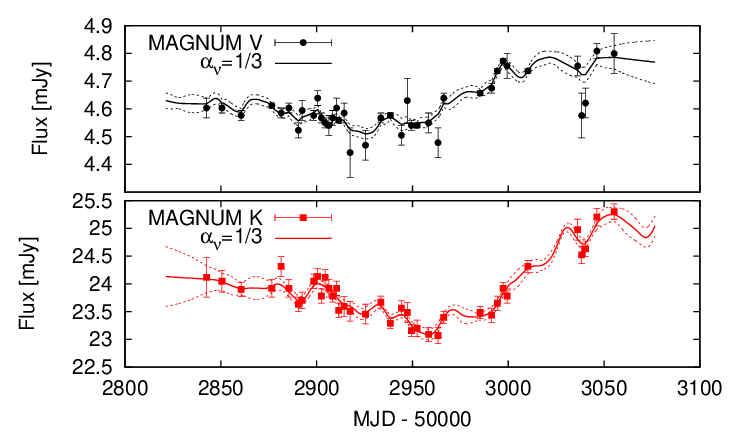}
}
\par \vspace{0.5cm}
\center{
\includegraphics[clip, width=3.2in]{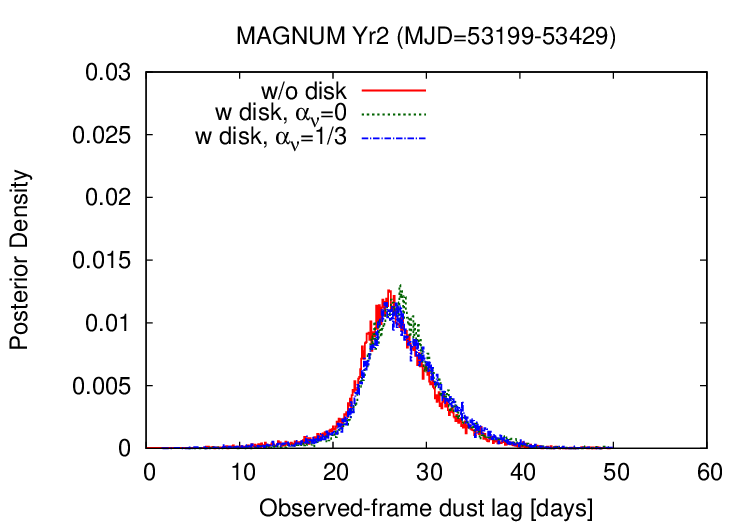}
\includegraphics[clip, width=3.4in]{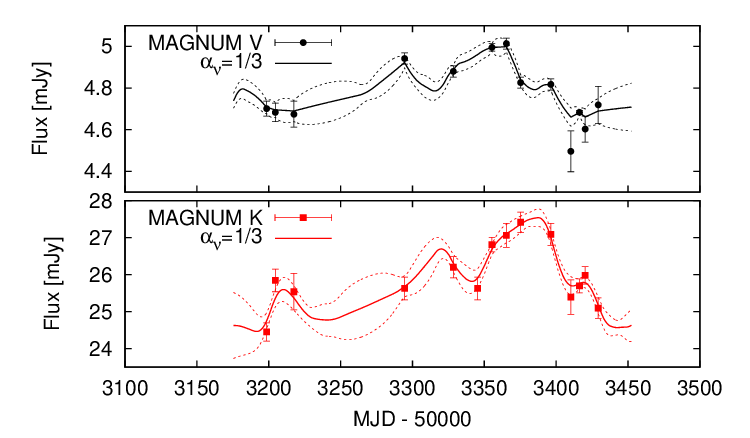}
}
 \caption{Left: the JAVELIN posterior distributions of the observed-frame dust lags derived from $V$- and $K$-band MAGNUM light curves at the first part (MAGNUM Yr1; top panel) and the second part (MAGNUM Yr2; bottom panel). 
 Different lines indicate different assumptions regarding the disk flux contribution to the $K$-band, where $\beta = f_{\nu}(K, \text{disk})/f_{\nu}(V, \text{disk}) = (\nu_{K}/\nu_{V})^{\alpha_{\nu}}$. Right: the MAGNUM $V$- and $K$-band light curves at the first (top) and second part (bottom). The Galactic extinction is not corrected. The damped random walk (DRW) model light curves for the $\alpha_{\nu}=1/3$ model are also shown with 1$\sigma$ uncertainties.}
 \label{fig:javelin_posterior}
\end{figure*}

We cross-check the results from the CCF analysis obtained above using another different lag estimation method.
Here we also consider the effects of the disk emission contribution to the $K$-band light curve on the lag estimation.

We use JAVELIN \citep{zu11,zu16} as a tool to derive the dust reverberation lags.
JAVELIN assumes a damped random walk (DRW) as a model for the AGN UV-optical continuum light curves, in which the DRW time series is defined by an exponential covariance function in the form of $\sigma^2 \exp(-\Delta t/\tau)$, where $\sigma$ is the asymptotic variability amplitude of the time series and $\Delta t$ is the time separating the two observations.
Then, a top-hat function is assumed as a transfer function $\Psi(t)$, and the responding light curve $l(t)$ is modelled using a driving light curve $c(t)$ as $l(t) = \int_{}^{} \Psi(t-t') c(t')dt'$ \citep{zu11,zu16}.
JAVELIN internally models and subtracts the mean of the light curve from the input light curve \citep[][]{zu11}; thus, the parameter estimations with JAVELIN are not affected by the flux contribution from the host galaxy emission to the light curves, as is the case for the CCF analysis.

The two-band photometric reverberation mapping function of JAVELIN can deal with a fractional contribution of the driving light curve to the responding light curve by introducing a parameter ($\beta$)\footnote{This parameter is referred to as $\alpha$ in \cite{zu16}, but we use the symbol $\beta$ instead to avoid confusion with $\alpha_{\nu}$.} defined as the ratio of the fluxes of the driving emission between the two bands  \citep{zu16}.
We utilise this function to examine the influence of the variable disk continuum emission in the $K$-band light curve on the lag analysis, as follows.

We take the MAGNUM $V$-band light curve as the driving disk light curve and the MAGNUM $K$-band light curve as the responding dust emission light curve, and we include the disk emission contribution to the $K$-band light in the JAVELIN modelling by fixing parameter $\beta$.
Note that here, the $V$-band light curve is assumed to be representative of the flux variation of the total UV-optical disk emission, which is driving the flux variation of the hot dust emission. This assumption is certified by observational facts that the AGN UV-optical disk emission generally shows well-correlated inter-band flux variation and that the empirical BLR/dust innermost radius$-$optical luminosity relationships are tight \citep[e.g.,][]{kor04,ben13,kos14,kok14,kok15,ese15}.

When the variable disk emission spectrum is modelled as a power-law in the form of $f_{\nu} \propto \nu^{\alpha_{\nu}}$, the ratio parameter $\beta$ can be expressed as follows [Eq.~(2) of \citet{kos14}]:
\begin{equation}
\beta = (\nu_{K}/\nu_{V})^{\alpha_{\nu}} = (\lambda_{V}/\lambda_{K})^{\alpha_{\nu}},
\end{equation}
where the effective wavelengths of the MAGNUM $V$-band and the $K$-band are $\lambda_{V} = c/\nu_{V} =  0.55~\mu\text{m}$ and $\lambda_{K} = c/\nu_{K} = 2.2~\mu\text{m}$, respectively \citep{yos14}.
It is uncertain whether the disk emission at the $K$-band can be estimated from the extrapolation of the optical power-law continuum, considering the possibility that the disk outer radius may be truncated due to the dominance of the disk's self-gravity \citep[e.g.,][and references therein]{goo03,kis08}.
We model the variable disk contribution to the $K$-band light curve by adopting the power-law indices between the $V$- and $K$-bands as $\alpha_{\nu}=\infty$, $0$, and $1/3$ ($\beta = 0$, 1, and 0.63, respectively; these values correspond to the assumptions of a disk truncation at the $K$-band (i.e. no disk contribution), a flatter disk spectrum than the thin disk model, and the thin disk model, respectively.
Currently, the two-band photometric reverberation mapping function of JAVELIN is the only way to deal with the disk flux contribution to the responding light curve self-consistently; however, as shown below, the disk flux contribution has only a minor effect on the lag estimate in the case of $K$-band dust reverberation mapping \citep[see also][]{kos14}.

Following the standard JAVELIN analysis procedure \citep{zu16}, first the DRW parameter ranges are constrained by analysing the full MAGNUM $V$-band light curves.
Then, the $V$- and $K$-band light curves at MAGNUM Yr1 and at Yr2, corresponding to significant cross-correlation signals (Fig.~\ref{fig:ccf_cccd}), are analysed separately by the two-band photometric reverberation mapping function of JAVELIN with the constrained DRW parameter ranges.
We adopt an upper limit of 50~days on the lag and 50~days on the width of the transfer function, and impose a minimum kernel width of 1 day to avoid solutions of the $\delta$ transfer function \citep[e.g.,][]{fau16,mud18}.

Fig.~\ref{fig:javelin_posterior} shows the posterior distribution of the dust reverberation lags measured from the MAGNUM light curves at MAGNUM Yr1 (upper panel) and at Yr2 (bottom panel), constructed from 100,000 Markov Chain Monte Carlo draws produced by JAVELIN; the median and $16-84$~\% percentiles of the lag distributions are summarised in Table~\ref{tbl-combinelag}.
The dust lag distributions derived by adopting different disk contribution models ($\alpha_{\nu} = \infty$, 0, and 1/3) are mostly consistent with each other, indicating that the disk contribution model has a negligible effect on the dust lag estimation \citep[see also][]{kos14}.
The analysed MAGNUM light curves, the best-fit DRW model light curves, and the 1$\sigma$ uncertainties in the case of the $\alpha_{\nu}=1/3$ model are also shown in the same figure.
The clearest lag signal of $32-34$~days can be seen in the $V$- and $K$-band light curves at MJD $\sim$ 52950 in the top right panel of Fig.~\ref{fig:javelin_posterior}, which is $\sim$ 1,000~days after the significant luminosity drop in Mrk~590.

The JAVELIN results of the dust reverberation lag estimation are consistent with the CCF result, although the uncertainties associated with the estimated lags are larger in the case of the cross-correlation analysis compared to the JAVELIN analysis.
The smaller estimation uncertainties in the JAVELIN analysis are probably due to the fact that JAVELIN assumes a specific time series model, i.e. the DRW model \citep[e.g.,][]{fau16}.

\subsection{Expected dust innermost radius from the AGN rest-frame $V$-band continuum luminosity}
\label{sec:expected_dust}

\begin{figure}
\center{
\includegraphics[clip, width=3.4in]{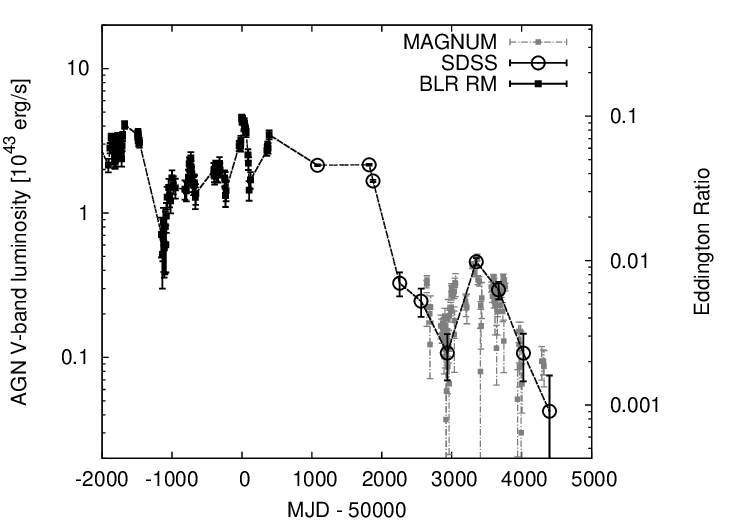}
}
 \caption{Long-term light curve of the AGN $V$-band luminosity of Mrk~590.
 The SDSS pseudo $V$-band AGN light curve is based on the observed $g$- and $r$-band light curves, and the BLR RM pseudo $V$-band AGN light curve is based on the observed spectroscopic continuum light curve (see the main text).
 The dotted line indicates log-linear interpolations between the data points. 
 For comparison, the AGN $V$-band luminosity derived from the observed MAGNUM $V$-band light curve (adopting $f_{V, \text{host}} = 4.396 \pm 0.015$~mJy; Section~\ref{sec:host_estimate_acs}) is also shown. 
 The right vertical axis shows the Eddington ratio calculated from the AGN $V$-band luminosity assuming $\alpha_{\nu}=0$, a bolometric correction of $L_{\text{bol}} = 9.26 \times L_{5100}$, and a black hole mass of $M_{BH} = 3.71 \times 10^{7} M_{\odot}$.
 }
 \label{fig:mrk590_transform}
\end{figure}

\begin{figure*}
\center{
\includegraphics[clip, width=6.4in]{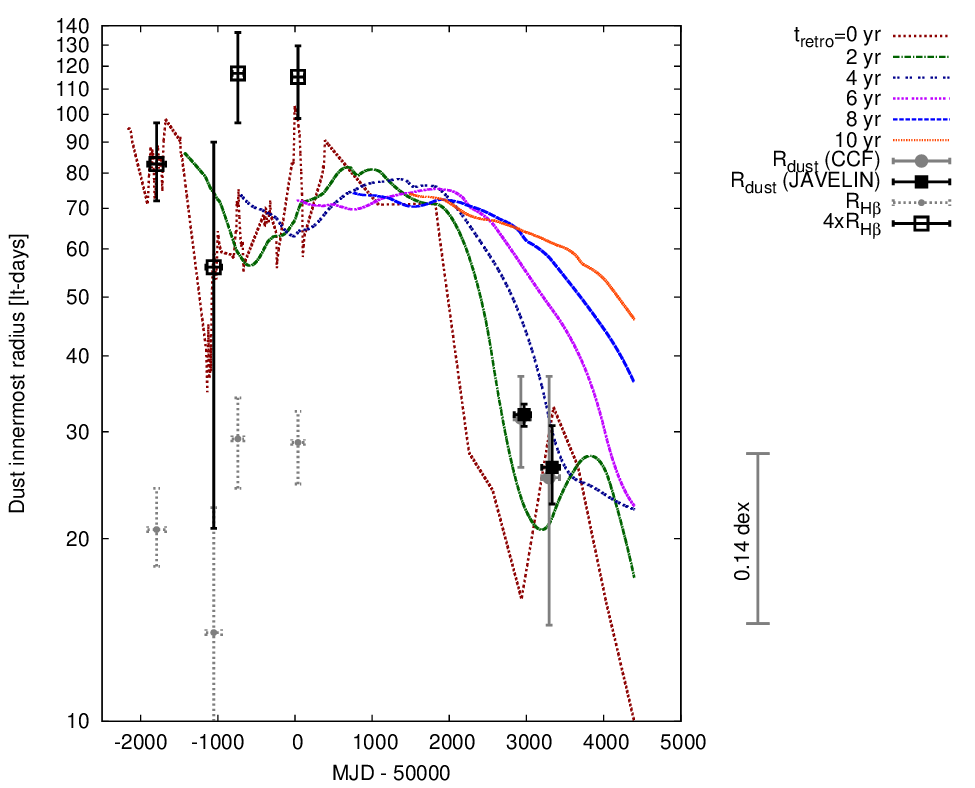}
}
 \caption{Dust reverberation measurements of the dust radius ($R_\text{dust} = c \times \tau_\text{rest, dust}$) as a function of time compared to the expected dust radius from the AGN $V$-band continuum light curve, based on the time-dependent dust radius$-$luminosity relation models [Eq.~(\ref{lag_lum_retro})]. The dust lag estimates of both methods, CCF/CCCD and JAVELIN ($\alpha_{\nu}=1/3$), are shown with slight horizontal offset for clarity. The error bars associated with the dust lag measurements are $16-84$\% percentiles, and the horizontal bars indicate the MJD range of the analysed light curves. The vertical bar at the right corner shows the intrinsic scatter of $\sigma_\text{add}=0.14$~dex of the empirical dust radius$-$luminosity relation. The $R_{\text{H}\beta~\text{BLR}}$ taken from Table~\ref{tbl-4}, and the expected dust radii at the epochs of $R_{\text{H}\beta~\text{BLR}}$ measurements, $R_{\text{dust}} = 4 \times R_{\text{H}\beta~\text{BLR}}$, are also shown.}
 \label{fig:mrk590_retroflux}
\end{figure*}

The empirical dust radius$-$luminosity relation for AGNs determined from the MAGNUM observations is derived against the $V$-band (5500~\AA) AGN luminosity \citep{sug06,kos14}; the dust radius $R_\text{dust}$ in parsecs can be expressed as a function of the rest-frame $V$-band AGN luminosity $L_{V}$, on the assumption of the accretion disk power-law index of $\alpha_{\nu}=0$, as follows \citep[Table~9 of][]{kos14}:
\begin{equation}
\log R_\text{dust} = -0.89 + 0.5 \log \left(\frac{L_{V}}{10^{44}~\text{erg/s}}\right)\ \ \ \left[ \sigma_\text{add} = 0.14~\text{dex} \right],
\label{lag_lum}
\end{equation}
where $\sigma_\text{add}$ is the intrinsic scatter in $\log R_\text{dust}$.

To infer the putative variations in the dust innermost radius along with the long-term optical variability of Mrk~590, we use the $g$- and $r$-band SDSS light curves to generate a pseudo $V$-band AGN light curve as follows.
First, the $g$- and $r$-band AGN light curves are derived by subtracting the host galaxy flux contribution given in Section~\ref{sec:host_estimate_sdss} ($3.373 \pm 0.029$~mJy and $7.538 \pm 0.047$~mJy) from the SDSS $g$- and $r$-band $\phi8''.3$ aperture flux light curves, respectively.
Considering the flat AGN disk continuum spectrum (i.e., $\alpha_{\nu} \sim 0$), the $V$-band [5500~\AA\ ($1+z$)] AGN light curve is then estimated by linearly interpolating between the $g$-band (4,686~\AA) and $r$-band (6,165~\AA) AGN light curves.
After the Galactic extinction correction ($A_V = 0.101$~mag), the $V$-band AGN light curve is converted to $V$-band luminosity using the luminosity distance of $d_L = 107$~Mpc (Section~\ref{sec:2}).
If we adopt the galaxy peculiar velocity uncertainty of 500~km~s$^{-1}$, as assumed by \cite{ben13}, the uncertainty of the luminosity distance of Mrk~590 is $7$~Mpc. 
This introduces 0.03~dex uncertainty on $\log R_\text{dust}$ that is negligible compared with the intrinsic scatter of 0.14~dex in the dust radius$-$luminosity relationship [Eq.~(\ref{lag_lum})]. In this work, we ignore the uncertainty in the luminosity distance of Mrk~590. Moreover, the 0.14~dex intrinsic scatter in Eq.~(\ref{lag_lum}) already includes the effects of the peculiar velocity uncertainties on the dust radius$-$luminosity relationship for 17 Seyfert galaxies.

As already described in Section~\ref{peterson_obs}, the spectroscopic 5100~\AA\ $(1+z)$ light curve from the H$\beta$ BLR reverberation mapping campaign by \cite{pet98b} is also converted to a pseudo $V$-band [5500~\AA\ $(1+z)$] AGN light curve by subtracting the host galaxy flux contribution and then applying a $K$-correction on the assumption of $\alpha_{\nu} = 0$ for the AGN optical continuum.
The host galaxy flux uncertainty described in Section~\ref{peterson_obs} is included.
The long-term AGN $V$-band light curve of Mrk~590 from the combination of SDSS data and H$\beta$ reverberation mapping monitoring data is shown in Fig.~\ref{fig:mrk590_transform}.
The MAGNUM $V$-band AGN light curve (Section~\ref{magnum_obs}, assuming $f_{V, \text{host}} = 4.396 \pm 0.015$~mJy) is also shown for comparison.
The $V$-band light curve estimated from the SDSS data agrees well with the MAGNUM light curve in the overlapping part.
For reference, the right vertical axis of Fig.~\ref{fig:mrk590_transform} shows the Eddington ratio calculated from the AGN $V$-band luminosity assuming $\alpha_{\nu}=0$, a bolometric correction of $L_{\text{bol}} = 9.26 \times L_{5100}$ \citep{ric06}, and a black hole mass of $M_{\text{BH}} = 3.71 \times 10^{7} M_{\odot}$ (Section~\ref{sec:2}).
We can see that the AGN luminosity drop from the 1990s to the 2000s corresponds to changes in the Eddington ratio from $\sim 0.05$ to $\sim 0.001$, consistent with the Eddington ratio variations from 0.061 (in 1984) to 0.006 (in 2015) estimated from the X-ray luminosity \citep[Table~5 of][]{koa16b}.
The amount of change in the Eddington ratio observed in Mrk~590 is very similar to that observed in another changing-look AGN Mrk~1018 \citep{mce16,hus16,nod18,dex19}, implying a common physical mechanism for their extreme disk emission variability.
By log-linear interpolation of the estimated $V$-band AGN light curve, we obtain a continuous $V$-band AGN luminosity time series, as shown in Fig.~\ref{fig:mrk590_transform}.

If we generalise Eq.~(\ref{lag_lum}) as a time-dependent relationship, there will be a time delay between the variations in AGN luminosity and the dust innermost radius \citep[e.g.,][]{kos09,pot10,kis13,sch17}.
\cite{kis13} model the variation in the innermost radius of the dust distribution in NGC~4151 determined from multi-epoch IR interferometric observations by assuming that the radius at a given epoch $t$ is set by the average AGN flux (``retro flux'') over the past $t_\text{retro}$ years, from $t-t_\text{retro}$ to $t$, which we denote as $<L_{V}>_{t-t_\text{retro}}^{t}$.
Using $t_\text{retro}$, the dust radius luminosity relation can be rewritten as
\begin{equation}
\log R_\text{dust}(t) = -0.89 + 0.5 \log \left(\frac{<L_{V}>_{t-t_\text{retro}}^{t}}{10^{44}~\text{erg}~\text{s}^{-1}}\right).
\label{lag_lum_retro}
\end{equation}
In Fig.~\ref{fig:mrk590_retroflux}, the dust reverberation measurements of the dust radius ($R_\text{dust} = c \times \tau_\text{rest, dust}$; Section~\ref{dust_lag_estimate}) as a function of time are compared with the expected dust radius based on Eq.~(\ref{lag_lum_retro}) for various $t_\text{retro}$ calculated using the interpolated $V$-band AGN continuum light curve in Fig.~\ref{fig:mrk590_transform}.
The luminosity uncertainty is not included in this model calculation, because it is much smaller than the intrinsic scatter of 0.14 dex in the dust radius$-$luminosity relationship [Eq.~(\ref{lag_lum})].
The larger values of $t_\text{retro}$ make the expected dust innermost radius larger at the epoch of the first detection of the dust reverberation lag, due to the delayed response of the dust radius to the significant luminosity drop at MJD $\simeq$ 52000.
From Fig.~\ref{fig:mrk590_retroflux}, we can conclude that the observed small dust reverberation radius requires a short delay time scale of 
\begin{equation}
t_\text{retro} \lesssim 4~\text{yr}.
\end{equation}
Also, the instantaneous response model ($t_{\text{resto}} = 0$~yr in Fig.~\ref{fig:mrk590_retroflux}) may be rejected, as the measured dust radius at MAGNUM Yr1 (MJD = $52843-53055$) is not as small as the expected dust radius predicted by this model.
The model parameter range of $t_\text{retro}$, which can best describe the observed dust radius, is $2~\text{yr} \lesssim t_\text{retro} \lesssim 4~\text{yr}$.
Note that this constraint on the dust replenishment time scale for Mrk~590 is not strongly affected by the integrated light curve history effect adopted in the retro-flux model [Eq.~(\ref{lag_lum_retro})], as the rapid, large luminosity decline event in $2000-2001$ almost solely dominates the overall change in the dust innermost radius; thus, a simple time-shift of the instantaneous response model also similarly constrains the dust replenishment time scale to less than $4~\text{yr}$.

\cite{kos09} monitored the change in dust reverberation lag in NGC~4151 and claim that the time scale of the replenishment of the dust distribution inside the innermost dust torus after the fade of UV-optical luminosity is as long as about 1~year, while \cite{kis13} claim that the variations in the dust innermost radius of NGC 4151 traced by the historic dust reverberation and IR interferometric measurements can be explained by adopting $t_{\text{retro}} \sim 6$~yr.
These results agree with our result that the dust distribution will not be replenished in the central region of AGNs immediately after the fade of the UV-optical luminosity. 
However, the constraint of $2~\text{yr} \lesssim t_{\text{retro}} \lesssim 4\text{yr}$ obtained for Mrk 590 is larger than the time scale obtained by \cite{kos09}, but smaller than that obtained by \cite{kis13} for NGC 4151.
This discrepancy may be attributed to the uncertainty in the estimates of \cite{kos09} and \cite{kis13}, owing to the limited dynamic range of the variations in luminosity and the inner radius of the dust torus observed in NGC 4151; the ratio of the highest to lowest fluxes of NGC 4151 is $\sim 4$.
Thus, only a two-fold larger dust innermost radius is predicted during the observed period, which is often buried under measurement errors of reverberation lag and interferometric measurements \citep[see also][]{sch15,sch17}.
Because we note a sudden drop in the AGN flux just before the dust reverberation measurements for Mrk 590, whose flux variation amplitude reaches approximately an order of magnitude, our constraints on $t_{\text{retro}}$ for Mrk 590 are considered to be more stringent than those for NGC 4151 in previous studies.

Another possible cause of the discrepancy in the dust replenishment time scale between Mrk~590 and NGC~4151 is intrinsic differences in the dust torus properties and rates of decline of the luminosity drop. To examine this possibility, further monitoring observations of Mrk~590, NGC~4151 and other AGNs are needed, which is beyond the scope of this paper.

\section{Discussion}

In this Section, we first show that the dust replenishment time scale of the innermost dust distribution in Mrk~590 is too short to be explained by the radial inflow of dust clouds, and then suggest that the dust replenishment is achieved by new dust grain formation in the BLR/innermost dust torus region (Section~\ref{sec:reform}).
Then, we consider the physical conditions of the BLR and accretion disk, and show that new dust grain formation is possible in these regions on several year time scales once the AGN luminosity drops (Section~\ref{sec:new_dust}).
In Section~\ref{sec:future}, we discuss how prospective observations of metal emission lines and NIR/mid-infrared (MIR) dust emission after rebrightening of Mrk~590 can further constrain the dust formation/destruction processes in the AGN.

\subsection{Reformation time scale of the innermost dust distribution}
\label{sec:reform}

The dust grain temperature in the AGN dust tori is determined by the radiative equilibrium between the thermal emission of the dust grains and energy input from the illumination of the AGN accretion disk emission \citep[e.g.,][]{bar87}.
The standard picture of the AGN dust torus is that the dust innermost radius corresponds to the dust sublimation radius defined by the dust grain temperature of $\sim 1,800$~K \citep[e.g.,][]{sug06,kos14,yos14}.
Therefore, it is naturally expected that the dust innermost radius in each individual AGN varies as a function of time according to the changes in AGN luminosity states, given as $R_{\text{dust}} \propto L_{\text{AGN}}^{0.5}$.
However, currently there are only a few observational constraints on the time scales of the destruction/reformation of the innermost dust distribution in AGNs.
From multiple $K$-band dust reverberation mapping and $K$-band interferometric observations for NGC~4151 ($R_{\rm dust} \sim 50$~lt-days), it has been suggested that the destruction/reformation time scale in NGC~4151 is in the range of $\sim 1-6$~yr; however, the interpretation of the observations is still under debate \citep[e.g.,][]{kos09,pot10,hon11,kis13,okn14}.

Our re-analysis of the MAGNUM observation described in  Section~\ref{dust_lag_estimate} shows that the dust radius had become as small as $R_\text{dust}(\text{faint}) \simeq 32$~lt-days by the period during $52,843-53,055$ (the year 2004), when the AGN was in the faint phase.
The $V$-band luminosity of Mrk~590 differs between the bright and faint phases by a factor of $\sim 7$ (Section~\ref{sec:expected_dust}), and from the empirical dust radius$-$luminosity relationship of $R_\text{dust} \propto L_{V}^{0.5}$ \citep{kos14}, the dust reverberation radius at the bright phase is expected to be $R_\text{dust}(\text{bright}) \simeq 7^{0.5} \times 32 = 85$~lt-days.
The observed BLR reverberation radius of Mrk~590 in its bright phase is $R_{\text{BLR, H}\beta} = 25.6^{+2.0}_{-2.3}$~lt-days \citep[Table~\ref{tbl-4};][]{pet98b,ben09,ben13}; thus, the ratio of the dust radius to the BLR radius is $R_\text{dust}(\text{bright})/R_{\text{BLR, H}\beta} \simeq 3.3$, which is consistent with the global $R_{\text{BLR, H}\beta}-R_\text{dust}$ relationship observed in other Seyfert galaxies, as shown in Fig.~\ref{fig:dust_blr}, within the intrinsic scatter of the relationship.
This consistency strongly suggests that the dust distribution between $R_\text{dust}(\text{faint})$ and $R_\text{dust}(\text{bright})$ had been replenished according to the AGN luminosity variations.

Additional supporting evidence for the receded torus innermost radius in Mrk~590 is the observed decrease in the $K$-band AGN flux.
As shown in Section~\ref{sec:nirdata_2mass_denis}, the ratio of the $K$-band AGN fluxes between the faint and bright phases is $2.32~\text{mJy}/20.44~\text{mJy}$.
If we assume that the $K$-band emission is from the black body surface at the torus inner wall with a fixed torus opening angle, the $K$-band luminosity $L_{K}$ is proportional to the area of the inner wall ($L_{K} \propto 4 \pi R_{\text{dust}}^2$); thus, the ratio of $R_{\text{dust}}$ between the faint and bright phase is $R_\text{dust}(\text{faint})/R_\text{dust}(\text{bright}) = (2.32~\text{mJy}/20.44~\text{mJy})^{0.5} \sim 0.3$.
This is in accordance with the reduction in the torus's innermost radius predicted from the optical luminosity light curve, as shown in Fig.~\ref{fig:mrk590_retroflux}.

SDSS Stripe~82 data reveal that the sudden decrease in the AGN luminosity of Mrk~590 occurred during the period from MJD=$51819-52288$ (in 2000-2001) (Section~\ref{sec:sdss_photometry}; Fig.~\ref{fig:mrk590_transform}).
As mentioned above, the replenishment of the dust distribution in the region between $R_\text{dust}(\text{bright}) \simeq 85$~lt-days and $R_\text{dust}(\text{faint}) \simeq 32$~lt-days takes place during the period from MJD $\simeq 52000$ (large luminosity drop) to MJD $\simeq 53000$ (dust lag measurement), i.e. it takes only 1,000~days (Section~\ref{dust_lag_estimate}).
Comparisons between the measured dust radius and the time-dependent radius$-$luminosity relation models given by Eq.~(\ref{lag_lum_retro}) provide a rough constraint on $t_\text{retro}$ as $t_\text{retro} < 4$~yr (Fig.~\ref{fig:mrk590_retroflux}).
From the considerations described above, the difference in sublimation radii between the bright and faint phases is expected to be
\begin{equation}
R_\text{dust}(\text{bright}) - R_{\text{dust}}(\text{faint}) \simeq 53~\text{lt-days}.
\label{eqn:sublimation_radii}
\end{equation}
If we assume that this adjustment is achieved by radial inflows of the dust clouds located outside of the former dust sublimation radius, Eq.~(\ref{eqn:sublimation_radii}) requires a high radial infall velocity of 
\begin{equation}
v_{r} = \frac{R_\text{dust}(\text{bright}) - R_{\text{dust}}(\text{faint})}{t_\text{retro}} \gtrsim 11,000~\text{km~s}^{-1}.
\end{equation}
For comparison, if we consider the extreme condition in which dust clouds freely fall from $R_\text{dust}(\text{bright})$ to $R_{\text{dust}}(\text{faint})$ due to the BH gravity just after the luminosity drop (i.e. the clouds have no angular momentum), the free fall velocity of the clouds at $R_{\text{dust}}(\text{faint})$ can be calculated as 
\begin{equation}
v_\text{free} = \left(\frac{2GM_\text{BH}}{R_\text{dust}(\text{faint})}-\frac{2GM_\text{BH}}{R_\text{dust}(\text{bright})}\right)^{1/2} \sim 2,700$~km~s${}^{-1},
\end{equation}
which is much lower than the required radial inflow velocity $v_{r}$, or equivalently, the corresponding free fall time scale ($t_{\text{free}} \sim 50$~yr) is much larger than the observed dust replenishment time scale $t_{\text{retro}}$.
These numbers suggest that the radial inflow scenario is unlikely \citep[see e.g.,][]{ste18,ros18}.

Instead, the replenishment can be achieved by rapid condensation of new dust grains after the dust sublimation radius moves inward following the AGN luminosity drop as $R_{\text{dust}} \propto L_{\text{AGN}}^{0.5}$.
Note that the dust sublimation radius denotes the hypothetical radius where the temperature of test dust particles (determined by the local radiative equilibrium between the irradiating AGN flux and dust thermal emission) equals the dust sublimation temperature; thus, it moves on a light-travel time scale following the AGN luminosity variations. The light-crossing time for the distance $R_{\text{dust}}(\text{bright}) - R_{\text{dust}}(\text{faint})$ [Eq.~(\ref{eqn:sublimation_radii})] is 53~\text{days}, which is much shorter than the observed replenishment time scale $2~\text{yr} \lesssim t_{\text{retro}} \lesssim 4~\text{yr}$.
\cite{kos09} also claim that the rapid variations in the dust reverberation lags of NGC~4151 are due to dust formation/destruction in the innermost region of the dust torus \citep[see also][]{ess19}.
Similarly, \cite{bar92} suggests that rapid dust reformation on a time scale of months following UV-optical luminosity variations is probably required to explain the multi-band NIR light curves of Fairall~9.

\subsection{Reformation mechanism of the innermost dust distribution}
\label{sec:new_dust}

To replenish the innermost dust torus within a time scale of $<4$~years, as observed in Mrk~590, new dust grain formation at $R_\text{dust} \sim 30$~lt-days is required just after the rapid decline in accretion disk luminosity.
In the following subsections, we consider two possible mechanisms to realise the rapid replenishment of the innermost dust distribution: new dust grain formation in the cooled BLR gas, and new dust grain formation in the cooled accretion disk atmosphere and its vertical inflation in the form of a failed dusty disk wind.

Before discussing the details of these dust grain formation scenarios, we summarise the general assumptions regarding the gas and dust in the AGN BLR and dust torus.
The dust sublimation temperatures of graphite and silicate grains are the temperatures at which the partial gas pressures of gas-phase C and Si equal the vapour pressure of graphite and silicate, respectively \citep[e.g.,][]{guh89,phi89,bas18}.
Assuming $Z=Z_{\odot}$ and a gas temperature of $T = 10,000$~K, the sublimation temperature of the graphite grain is approximately $T_\text{sub}(\text{K}) = 81,200/(66.003 - \ln n) \sim 1,800-2,000$~K, where $n$ is the total gas density; here we assume $n \sim 10^{10}-10^{11}~\text{cm}^{-3}$ for the BLR gas clouds \citep{bas18}.
The sublimation temperature of the graphite grains is higher by $\sim 300-500$~K compared to silicate grains; additionally, larger dust grains have lower temperatures for a given UV radiation field, due to more efficient radiative cooling compared with smaller dust grains.
Thus, the large graphite grains are expected to play an important role in controlling the innermost radius of AGN dust tori \citep[][]{mor12,yos14,vel16,hon17b,bas18}.
The theoretically expected dust sublimation temperature of $T_\text{sub} \sim 1,800-2,000$~K is largely consistent with the upper limit on the dust temperature at the innermost region of the AGN dust tori observationally inferred from the colour temperature of the NIR variable continuum emission SED \citep{tom06,yos14}.
Although there is a 4-fold difference between the mean reverberation radii of the innermost dust torus and H$\beta$ BLR, the presence of broad/intermediate velocity width emission lines of elements with various ionisation potentials suggests that the BLR gas clouds and the dust clouds are contiguously distributed, in which the dust sublimation radius defines the outermost radius of the BLR. \citep{lao93,kos14,bas18}.

\subsubsection{New dust formation in the radiatively cooled BLR gas}
\label{sec:dust_in_blr}

In Fig.~\ref{fig:mrk590_retroflux}, the dust reverberation radius in the faint phase of Mrk~590 is coincident with the H$\beta$ BLR reverberation radius in the bright phase.
Assuming the contiguously distributed BLR and dust torus structure, it is natural to consider that the new dust grain formation occurs in the region that was once the dust-free BLR gas after the rapid decline in the accretion disk luminosity.
We can consider that the opening angle of the newly formed dust innermost distribution is determined by the balance between the gravitational force and radiative pressure, and becomes slightly smaller in the faint state than in the bright state \citep[see][]{bas18}.

Once the AGN radiation input decreases and the BLR gas cools down, the BLR gas clouds can potentially be a suitable site for new dust formation given their high density \citep[see e.g.,][]{phi89,elv02,mai06}.
Consider the BLR gas condition just inside the dust sublimation radius in the bright phase of Mrk~590, and the new dust formation of carbon (C) grains after the AGN luminosity drop.
When the AGN accretion disk luminosity (which is proportional to the ionisation parameter) decreases, the BLR gas begins to cool rapidly from the initial gas temperature of $\sim 10,000$~K to $\lesssim 2,000$~K via metal line cooling over a period of $\lesssim 20$~days,  due to the high density  \citep[$n\sim 10^{10}-10^{11}$~cm${}^{-3}$; e.g.,][]{nam16,ich17,sar18}.
The radiatively cooled BLR gas can then initiate dust grain formation, as the partial gas pressure of C becomes larger than the vapour pressure of the bulk condensate at the gas temperature of $\lesssim 2,000$~K, i.e. the gas shifts to a supersaturated state \citep[e.g.,][]{sal77,phi89,elv02,noz13,bas18}.

Assuming a sticking probability of unity, the dust formation time scale can be estimated approximately as
\begin{eqnarray}
t_{\text{form}} &\simeq& \frac{4}{3}\pi a^{3} \rho_\text{bulk} \left[ 4\pi a^2 \sqrt{\frac{m_{C}}{2\pi k_B T}} P_{C} \right]^{-1},\nonumber\\
&\sim& 26~\text{days}~\left( \frac{a}{0.1~\mu \text{m}} \right) \left( \frac{T}{2,000~\text{K}} \right)^{-1/2} \left( \frac{n}{10^{10}~\text{cm}^{3}} \right)^{-1}
\label{eqn:t_form}
\end{eqnarray}
where $a$ is the radius of a dust grain, $\rho_\text{bulk}=2.26$~g~$\text{cm}^{-3}$ is the bulk mass density of graphite and $m_{C}$ is the mass of a C atom.
The partial gas pressure of C is $P_{C} = n_{C} k_{B} T$, where $T$ is the temperature of gas-phase C and $n_{C}$ is the number density of C atoms in the gas \citep[][]{phi89,kis13,sch17}.
Here $n_{C} = 10^{-3.44} n$ (i.e. $Z=Z_{\odot}$) is assumed; the higher metallicity of $Z > Z_{\odot}$ leads to more rapid dust grain formation.
The newly formed dust and gas can have different temperatures because the collisional coupling between the dust grains and gas particles is very weak, and the dust grains are expected to achieve radiative equilibrium between the AGN radiation input and radiative cooling of the dust grains effectively instantaneously \citep[see e.g.,][]{phi89,nen08a,vel16,ich17}.
Eq.~(\ref{eqn:t_form}) indicates that the new dust formation in the BLR gas clouds can produce large graphite grains of $a \lesssim 0.1-1~\mu$m within a time scale of $< 1$~year.

A potential complication of this picture is that the BLR gas continues to be exposed to AGN radiation from the inner part of the accretion disk, and the BLR gas is probably not fully neutralised.
The dust sublimation temperature in the photoionised gas is lower than that assumed above; thus, the dust grain formation is not as efficient as that calculated above \citep[e.g.,][]{guh89,der17,bas18}.
Moreover, if we assume that the BLR gas density is determined by the incident radiation pressure compression \citep[e.g.,][]{ste14,bas18}, the BLR gas density in the faint phase is lowered by the same factor as the luminosity decrease.
Detailed theoretical calculations, including of changes in the BLR gas density and ionisation states of the gas-phase species, chemical reactions, and formation of seed clusters of molecules and their growth \citep[e.g.,][]{noz13,der17}, are required to fully understand the extent of dust grain formation in the AGN BLR gas clouds and the size distribution that should be expected for the newly formed dust, which is beyond the scope of this paper.

\subsubsection{New dust formation in the cooled accretion disk atmosphere and dust torus reformation by failed dusty disk wind}
\label{sec:dust_in_disk}

As assumed in the failed radiatively accelerated dusty outflow model discussed in \cite{bas18} \citep[originally proposed by][]{cze11,cze17}, the accretion disk atmosphere, where the effective temperature is below 2,000~K, could be a relevant place for new dust grain formation.
As noted above, dust grain formation occurs when the gas is neutral \citep[e.g.,][]{der17,bas18}; the accretion disk atmosphere is probably a better place for dust grain formation, compared to the BLR gas, as the equatorial plane of the disk is not directly exposed by the inner accretion disk radiation.

The luminosity decrease of Mrk~590 is interpreted as an intrinsic change in the mass accretion rate of the accretion disk \citep{den14}; thus, it is natural to consider that the accretion disk radius at the effective temperature of $T_\text{eff}=2,000$~K \citep[referred to as $R_\text{in}$ in][]{bas18} had moved inward in the faint phase of Mrk~590.
According to Eq.~(\ref{eqn:t_form}), the new dust grain formation in the disk atmosphere at $R \gtrsim R_\text{in}$ (with $T_\text{eff} \lesssim$~2,000~K) occurs effectively instantaneously, as the gas density of the disk atmosphere is expected to be much denser than the typical BLR gas density of $\sim 10^{10}-10^{11}~\text{cm}^{-3}$ \citep{jia19}.

As discussed by \cite{bas18}, the high temperature dust grains in the disk atmosphere provide a sufficiently large opacity of $\kappa \sim 10$~cm${}^{2}~g{}^{-1}$ to vertically puff-up the disk height by radiation pressure from the local accretion disk IR emission \citep{cze11,cze17,ess19,che19}.
The vertical motion of the puffed-up region is governed by the balance between the radiation pressure and the SMBH gravity acting on the gas/dust particles, which forms a failed dusty disk wind \citep{cze17}.
Then, the puffed-up region is exposed to radiation from the inner accretion disk, and eventually forms the dynamical gas/dust structures responsible for the BLR and innermost dust torus region, which are divided by the dust sublimation radius of graphite grains of $a \sim 0.1$~$\mu$m determined by the irradiating accretion disk luminosity.
The motion of the failed dusty disk wind at a given $R$ ($>R_\text{in}$) is approximately described as oscillations between the accretion disk surface and a peak height with the local Keplerian period, $\Omega_{K} = \sqrt{GM_\text{BH}/R^3}$ \citep{cze17}.

In the case of Mrk~590, we can assume a situation where the dusty disk wind launched from the accretion disk surface at $R_{\text{dust}}$, which is determined by the MAGNUM observations as $R_{\text{dust}} \sim 30$~lt-days\footnote{On the assumption of the standard disk model of \cite{sha73}, the disk temperature at $R_\text{dust}=30$~lt-days is $T_\text{eff} \simeq 275$~K $\times \left( M_\text{BH}/3.71 \times 10^{7}M_{\odot} \right)^{1/2} \times \left[ \left(L_\text{bol}/L_\text{Edd}\right)/0.005 \right]^{1/4} \times \left( \epsilon/0.1 \right)^{-1/4} \times \left( R_\text{dust}/30~\text{lt-days} \right)^{-3/4}$ [Eq.~(6) of \cite{bas18}], which is well below the dust sublimation temperature.}, reaches the peak height on a time scale of
\begin{eqnarray}
t_\text{wind} = \frac{2 \pi}{4\Omega_{K}} &=& \frac{\pi}{2}\sqrt{\frac{R_{\text{dust}}^3}{GM_\text{BH}}}\nonumber\\
&\simeq& 15~\text{yr} \left( \frac{R_{\text{dust}}}{30~\text{lt-days}} \right)^{3/2} \left( \frac{M_\text{BH}}{3.71 \times 10^{7}M_{\odot}} \right)^{-1/2}.
\label{eqn:disk_wind}
\end{eqnarray}
Given that the dust emission can emerge before the dusty disk wind reaches the peak height, Eq.~(\ref{eqn:disk_wind}) gives an upper limit on the dust replenishment time scale after the luminosity decline.
This time scale is consistent with the observed time scale $t_{\text{retro}}$ ($2~\text{yr} \lesssim t_\text{retro} \lesssim 4~\text{yr}$; Section~\ref{sec:expected_dust}), implying that the failed dusty disk wind launched from the cooled accretion disk atmosphere in the faint phase of Mrk~590 can explain the replenishment of the innermost dust distribution within a few years after the luminosity decline in 2002$-$2003.

\subsection{Possible destruction/outflow of the newly formed dust after AGN rebrightening}
\label{sec:future}

It is possible that Mrk~590 may experience rebrightening of the AGN luminosity sometime in the future.
Actually, Chandra X-ray and Hubble Space Telescope (HST) UV observations from 2014 \citep{mat18} and the Very Large Telescope/MUSE observations in 2017 \citep{rai19} hint at a reawakening of AGN activity in Mrk~590.
If so, the dust grains that formed in the innermost region during the faint phase will sublimate rapidly due to the enhanced vapour pressure of the dust grains heated by the re-brightened AGN luminosity \citep[see e.g.,][]{wax00,hon11,kis13,jia17b,bas18}.
The time scale of the dust destruction due to increased AGN luminosity may be shorter than, or of the same order as, the time scale of the dust formation \citep[][]{vel16,sch17,bas18}.
Therefore, when a sudden faint-to-bright luminosity transition occurs in Mrk~590, we will observe an increase in the dust reverberation radius on a time scale less of than $t_\text{retro}$, as determined for Mrk~590 in this work  [Eq.~(\ref{eqn:t_form})].
Metal-rich gas released from the evaporated dust may produce transient metal emission lines (such as narrow Fe~II lines) at UV-optical wavelengths \citep[e.g.,][]{jia17b}.

After AGN re-brightening, a portion of the dust grains that formed during the faint phase will be ejected from the dust torus as a radiatively-driven dusty outflow \citep{nam16,hon17b,wil19,taz19}.
The dusty outflow launched from the inner part of the dust torus is suggested to be responsible for the parsec-scale MIR ($8-13$ $\mu$m) dust emission originating from the polar dusty region recently discovered in local AGNs (including Mrk~590) by IR high-resolution imaging observations \citep[e.g.,][]{hon13,lop16,asm16}.
The rapid AGN re-brightening, accompanied by the larger UV-optical  emission radius of the accretion disk, can temporarily increase the radiation pressure on the inner dust torus along the vertical direction to the disk plane, which enables launch of the dusty outflow towards the polar region.
Therefore, the AGN fading-brightening cycle, as observed in Mrk~590, may provide a channel to produce the additional polar dust structure around the AGN and the corresponding MIR dust emission component.

Future dust reverberation mapping and UV-optical spectroscopic observations over the faint-to-bright luminosity transition will not only enable tracing of the rapid increase in the dust innermost radius in Mrk~590 but should also reveal the potentially emergent new dust emission component in the MIR wavelength range.

\section{Conclusions}

To test the hypothesis that the unexpectedly small dust reverberation radius in Mrk~590 (Fig.~\ref{fig:dust_blr}) is due to the large luminosity decrease that occurred before the dust reverberation measurement obtained by the MAGNUM project in 2003-2007, we have examined the long-term optical/NIR flux variability of Mrk~590 by combining the SDSS Stripe 82 light curve data obtained in 1998$-$2007 and MAGNUM data.
The SDSS $g$- and $r$-band light curves have revealed that the optical emission of Mrk~590 decreased significantly over a very short time scale during 2000 and 2001, which marks the start of the ``changing-look'' phenomenon in this AGN.
A large luminosity decrease has also been discovered in the $K$-band dust torus emission between 1998 (2MASS, DENIS) and 2003 (MAGNUM).

The combination of SDSS and MAGNUM data indicates that $V$- and $K$-band observations by the MAGNUM project were carried out in the faint phase of Mrk~590.
We have reanalyzed the dust reverberation lag of Mrk~590 by dividing the MAGNUM light curves into two parts (MAGNUM Yr1 and Yr2, corresponding to MJD = $52642-53055$ and MJD = $53199-54321$, respectively), and have shown that the $R_\text{dust}$ had already become small ($\simeq 32$~lt-days) by the year 2004 according to the significant AGN luminosity drop during 2000 and 2001.
These observations suggest that the innermost dust distribution in Mrk~590 had been rapidly replenished on a time scale of $< 4$ years.
This time scale is too short for the radial inflow of the dust clouds to replenish the innermost dust distribution between the dust sublimation radii expected at the bright and faint luminosity states of Mrk~590 [Eq.~(\ref{eqn:sublimation_radii})].
Instead, we have proposed that the dust replenishment in Mrk~590 had been achieved either by new dust formation in the radiatively-cooled BLR gas clouds or by the dusty disk wind of the newly formed dust in the cooled accretion disk atmosphere \citep[based on the failed radiatively accelerated dusty outflow model;][]{cze11,cze17,bas18}.

If the AGN of Mrk~590 rebrightens, the newly formed dust will be rapidly destroyed, and the dust innermost radius will be set to the dust sublimation radius determined by the re-brightened AGN luminosity.
It may also be possible that a part of the dust is radiatively  accelerated and ejected from the dust torus to form the dusty outflow, which may produce an additional MIR dust emission component.
Future multi-wavelength monitoring observations for Mrk~590 will allow continuous dust destruction/formation phenomena at the innermost region of the dust torus to be traced.

\section*{Acknowledgements}

This work was supported by JSPS KAKENHI Grant Number 17J01884.
We thank Ari Laor, Ryo Tazaki, Kohei Ichikawa, and Hirofumi Noda for valuable discussion.

This publication makes use of data products from the Two Micron All Sky Survey, which is a joint project of the University of Massachusetts and the Infrared Processing and Analysis Center/California Institute of Technology, funded by the National Aeronautics and Space Administration and the National Science Foundation.

The DENIS project has been partly funded by the SCIENCE and the HCM plans of the European Commission under grants CT920791 and CT940627.
It is supported by INSU, MEN and CNRS in France, by the State of Baden-W\"urttemberg in Germany, by DGICYT in Spain, by CNR in Italy, by FFwFBWF in Austria, by FAPESP in Brazil, by OTKA grants F-4239 and F-013990 in Hungary, and by the ESO C\&EE grant A-04-046.
Jean Claude Renault from IAP was the Project manager.  Observations were  
carried out thanks to the contribution of numerous students and young 
scientists from all involved institutes, under the supervision of  P. Fouqu\'e, survey astronomer resident in Chile.

Funding for the SDSS and SDSS-II has been provided by the Alfred P. Sloan Foundation, the Participating Institutions, the National Science Foundation, the U.S. Department of Energy, the National Aeronautics and Space Administration, the Japanese Monbukagakusho, the Max Planck Society, and the Higher Education Funding Council for England. The SDSS Web Site is http://www.sdss.org/.

The SDSS is managed by the Astrophysical Research Consortium for the Participating Institutions. The Participating Institutions are the American Museum of Natural History, Astrophysical Institute Potsdam, University of Basel, University of Cambridge, Case Western Reserve University, University of Chicago, Drexel University, Fermilab, the Institute for Advanced Study, the Japan Participation Group, Johns Hopkins University, the Joint Institute for Nuclear Astrophysics, the Kavli Institute for Particle Astrophysics and Cosmology, the Korean Scientist Group, the Chinese Academy of Sciences (LAMOST), Los Alamos National Laboratory, the Max-Planck-Institute for Astronomy (MPIA), the Max-Planck-Institute for Astrophysics (MPA), New Mexico State University, Ohio State University, University of Pittsburgh, University of Portsmouth, Princeton University, the United States Naval Observatory, and the University of Washington.

This work is partly based on observations made with the NASA/ESA Hubble Space Telescope, obtained from the data archive at the Space Telescope Science Institute. STScI is operated by the Association of Universities for Research in Astronomy, Inc. under NASA contract NAS 5-26555.

%%%%%%%%%%%%%%%%%%%%%%%%%%%%%%%%%%%%%%%%%%%%%%%%%%

%%%%%%%%%%%%%%%%% APPENDICES %%%%%%%%%%%%%%%%%%%%%

\appendix

\section{Host galaxy flux estimation}
\label{sec:host_estimate}

Here we estimate the $V$-band (Section~\ref{sec:host_estimate_acs}) and $g$- and $r$-band (Section~\ref{sec:host_estimate_sdss}) host galaxy flux contributions to the $\phi 8''.3$ circular aperture, performing GALFIT AGN/host decomposition of a HST/ACS F550M-band image and SDSS Stripe82 deep coadded $g$- and $r$-band images of Mrk~590, respectively.
In the main text, the estimated host galaxy fluxes are subtracted from the $\phi 8''.3$ circular aperture light curves of Mrk~590 to derive the host-subtracted AGN light curve.

\subsection{GALFIT modelling for the HST/ACS HRC F550M image}
\label{sec:host_estimate_acs}

\begin{figure*}
\center{
\includegraphics[clip, width=5.4in]{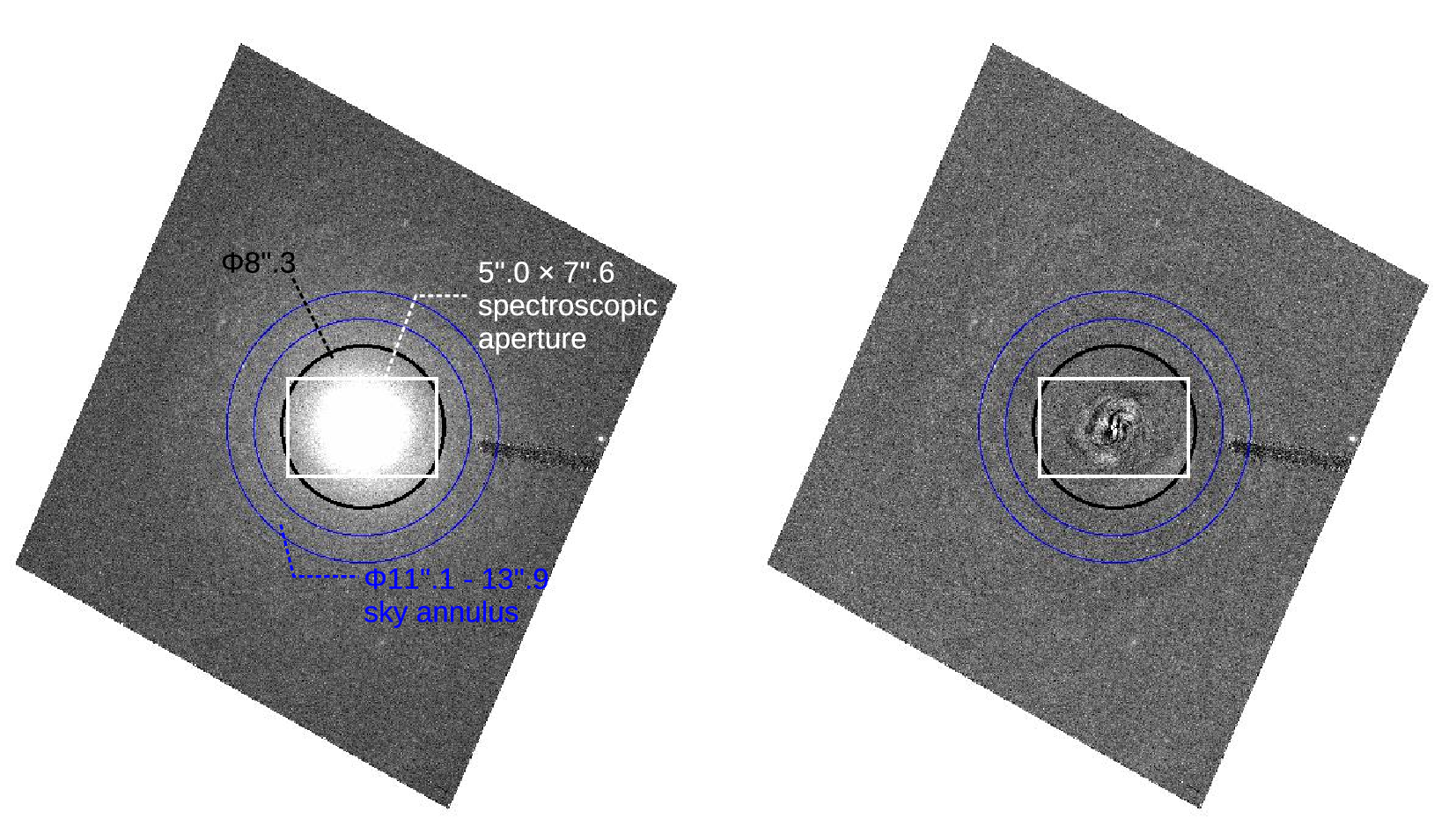}
}
 \caption{Left: HST/ACS HRC F550M image of Mrk~590. North is up and east is left. Right: GALFIT best-fit model-subtracted image. White rectangular region indicates the $7''.6 \times 5''$ spectroscopic aperture of Peterson et al. (1998) and the black circle indicates the MAGNUM $\phi 8''.3$ aperture. The blue annulus is the $\phi 11''.1-13''.9$ sky annulus aperture adopted by the MAGNUM photometry.}
 \label{fig:GALFIT}
\end{figure*}

\begin{table*}
\centering
\footnotesize
%\scriptsize
\caption{GALFIT modelling of the HST/ACS HRC F550M image of Mrk~590}
\label{tbl-galfit}
\begin{tabular}{c c c c c c c}
\hline
(1)   & (2)                &     (3)            & (4)     & (5) & (6)   & (7)\\
Model & Sky                & $m_{\text{stmag}}$ & $R_{e}$ & $n$ & $b/a$ & Note\\
      & (count~s${}^{-1}$) &                    & (kpc)   &     &       & \\
\hline
$\rm{S\acute{e}rsic}$ fixed  & $6.774 \times 10^{-3}$ &  $17.9888$ & \dots & \dots &  \dots & PSF \\
 &  &  $16.3669$ & $0.4026$ &  [$1.22$] & 0.6186 & Inner bulge \\
 &  &  $15.7723$ & $0.6912$ &  [$0.59$] & 0.9713 & Bulge \\
 &  &  $14.2761$ & $2.8854$ &  [$1$]    & 0.9056 & Disk \\\hline
\end{tabular}
\\
\begin{flushleft}
In the ``$\rm{S\acute{e}rsic}$ fixed'' model, the $\rm{S\acute{e}rsic}$ indices $n$ given in square brackets are fixed to the values given in \cite{ben09}. The zero ST magnitude is defined as $m_{\text{stmag},0} = -2.5   \log(\text{Photflam}) - 21.10 = 24.488$~mag. The pixel scale is $0''.025$~pixel${}^{-1}$, and the angular scale of 0.495 kpc~$\text{arcsec}^{-1}$ is assumed.
\end{flushleft}
\end{table*}

\cite{ben06,ben09,ben13} estimate the host galaxy flux contribution to the $5''.0 \times 7''.6$ rectangular spectroscopic aperture adopted for the H$\beta$ reverberation mapping observation data of \cite{pet98b}, using the HST/Advanced Camera for Surveys (ACS) High Resolution Channel (HRC) F550M filter image for Mrk~590, obtained on 2003 December 18 ($\lambda_{c} = 5,580$~\AA\ and $\Delta\lambda = 547$~\AA), as follows:
By modelling the AGN + host galaxy using a two-dimensional fitting algorithm GALFIT \citep{pen11}, the AGN-subtracted $5''.0 \times 7''.6$ aperture flux is calculated as $f_{\lambda, \text{F550M}} = 4.676 \times 10^{-15}$~erg~s${}^{-1}$~cm${}^{-2}$~\AA${}^{-1}$ \citep{ben13}.
A colour correction factor $f_{\lambda, (1+z)\text{5100~\AA}}/f_{\lambda, \text{F550M}}$ is derived to be $0.848$ using the bulge template spectrum of \cite{kin96}; thus, the host galaxy flux at $\lambda_{\text{obs}} = (1+z) \text{5100~\AA}$ is $f_{\lambda, (1+z)\text{5100~\AA}} = 3.965 \times 10^{-15}$~erg~s${}^{-1}$~cm${}^{-2}$~\AA${}^{-1}$ \citep{ben09,ben13}.

Using the same HST/ARC image, here we re-evaluate the host galaxy GALFIT modelling.
The HST/ACS HRC F550M observation for Mrk~590 consists of three exposures with exposure times of 120, 300, and 600 s (total exposure time: 1,020 s), each of which is split into two equal subexposures ({\tt CR-SPLIT}) to facilitate the rejection of cosmic rays.
Even the longest exposure images do not suffer from saturation, because the AGN component of Mrk~590 was in the faint phase at the epoch of the HST observation \citep{ben06}.

Pipeline products of the HST/ACS HRC F550M images ({\tt flt} images) were downloaded from the Mikulski Archive for Space Telescopes (MAST).
These images are combined into a final distortion-corrected drizzle-combined image using {\tt AstroDrizzle}\footnote{\href{https://drizzlepac.readthedocs.io/en/deployment/index.html}{https://drizzlepac.readthedocs.io/en/deployment/index.html}} adopting default settings, with the exception of sky subtraction.
A noise image for the drizzle-combined image is created using {\tt AstroDrizzle} again but adopting {\tt final\_wht\_type = ERR}.
Moreover, a model PSF image at the position of the AGN is created using a web interface of {\tt TinyTim} \citep{kri11}.

The image decomposition for the HST/ACS HRC F550M image of Mrk~590 is performed using GALFIT.
Following the analysis of \cite{ben09}, a PSF (AGN), three $\rm{S\acute{e}rsic}$ components (host galaxy) and a sky background are used as model components to be fitted.
Because the HST/ACS HRC has a limited field-of-view of $29''.1 \times 25''.4$ (Fig.~\ref{fig:GALFIT}), the sky background estimation is probably affected by the unmodelled extended host galaxy component \citep[][and see Section~\ref{sec:host_estimate_sdss}]{ben09}.
The GALFIT fitting is performed adopting fixed values of $\rm{S\acute{e}rsic}$ indices of $n = $1.22, 0.59, and 1.00 for the inner bulge, bulge, and disk components, respectively, as derived by \cite{ben09}.
The best-fitting parameters are listed in Table~\ref{tbl-galfit} (``$\rm{S\acute{e}rsic}$ fixed'' model).
From this fitting, we recover best-fitting parameters that are consistent with those derived by \cite{ben09}.

By subtracting the best-fitting PSF component and sky background component from the input image, we derive the $\phi 8''.3$ circular aperture host galaxy flux as $f_{\lambda, \text{F550M}} = 5.362^{+0.015}_{-0.015} \times 10^{-15}$~erg~s${}^{-1}$~cm${}^{-2}$~\AA${}^{-1}$ or $f_{\nu, \text{F550M}} = 5.569^{+0.015}_{-0.015}$~mJy, where the up-to-date inverse sensitivity calibration value is adopted ({\tt Photflam}$ = 5.817\times 10^{-19}$~erg~s${}^{-1}$~cm${}^{-2}$~\AA${}^{-1}$~count${}^{-1}$).
Here the flux uncertainty is conservatively evaluated by adding uncertainties from the noise map and GALFIT modelling of all the components in quadrature.
The $\phi 8''.3$ circular aperture host+AGN flux of the sky-subtracted, PSF-unsubtracted image is $f_{\nu, \text{F550M}}(\text{total}) = 5.814$~mJy; thus, the AGN flux at the epoch of the HST observation (2003 December 18; MJD = $52991.1$) is $f_{\nu, \text{F550M}}(\text{AGN}) = (5.814 - 5.569)$~mJy $= 0.245^{+0.015}_{-0.015}$~mJy\footnote{\cite{ben09} estimate the HST/ACS FRC F550M count flux of the AGN component as $f_{\nu, \text{F550M}}(\text{AGN}) = $419.6 count~s${}^{-1}$. Using {\tt Photflam}$ = 5.817\times 10^{-19}$~erg~s${}^{-1}$~cm${}^{-2}$~\AA${}^{-1}$~count${}^{-1}$, it can be converted to $f_{\nu, \text{F550M}}(\text{AGN}) = 0.253$~mJy, being consistent with our estimate of $f_{\nu, \text{F550M}}(\text{AGN}) = 0.245^{+0.015}_{-0.015}$~mJy.}.
By assuming the AGN optical continuum power-law index as $\alpha_{\nu} = 0$\footnote{Because there is no accurate measurement of the AGN disk continuum spectrum for Mrk~590, we assume a power-law spectral shape with $\alpha_{\nu} = 0$ throughout this work unless otherwise stated.}, this AGN flux can be converted into monochromatic AGN luminosity at $\lambda_{\text{rest}} = 5100$~\AA\ as $L_{V}(\text{AGN}) = 0.180^{+0.011}_{-0.011} \times 10^{43}$~erg~s$^{-1}$, being consistent with SDSS or MAGNUM photometry (Fig.~\ref{fig:mrk590_transform}) within their $2\sigma$ errors.
As a consistency check, the $5''.0 \times 7''.6$ rectangular spectroscopic aperture flux is measured for our best-fitting sky-subtracted, PSF-subtracted GALFIT model to be $f_{\lambda, \text{F550M}} = 4.676 \times 10^{-15}$~erg~s${}^{-1}$~cm${}^{-2}$~\AA${}^{-1}$, which is consistent with the value derived by \cite{ben13}.

The above estimate of the $\phi 8''.3$ circular aperture $V$-band flux is not directly comparable to the host galaxy flux contribution estimated by \cite{kos14} ($f_{\nu, V} = 4.24 \pm 0.06$~mJy).
This is because in the latter case, the host galaxy flux through the $\phi 11''.1-13''.9$ annulus aperture (normalised to the area of the $\phi 8''.3$ circular aperture) is further subtracted to match the standard MAGNUM photometry, where the sky reference aperture of the $\phi 11''.1-13''.9$ ring is used \citep[Fig.~\ref{fig:GALFIT}; see][]{sug06,sak10,kos14}.
To check the consistency between the HST/ACS GALFIT modelling result and that of \cite{kos14}, first we measure the $\phi 11''.1-13''.9$ annulus aperture flux (corrected for the area difference between the $\phi 11''.1-13''.9$ annulus (54.98~arcsec${}^2$) and $\phi 8''.3$ circular aperture (54.11~arcsec${}^2$) of the HST/ACS HRC image as $f_{\nu, \text{F550M}, \text{annulus}} = 0.905^{+0.006}_{-0.006}$ mJy.
Thus, the host galaxy flux contribution to the standard MAGNUM aperture is $f_{\nu, \text{F550M}} = (5.569 - 0.905)$~mJy = $4.664^{+0.016}_{-0.016}$~mJy.
Using the bulge template spectrum of \cite{kin96} (redshifted by $z=0.02639$) and the Galactic extinction coefficient of \cite{sch11} to calculate the colour correction term ($f_{\nu, V}/f_{\nu, \text{F550M}}=0.9425$), we obtain $f_{\nu, V} = 4.396^{+0.015}_{-0.015}$~mJy.
This estimate is consistent with that given in \cite{kos14} within their 3$\sigma$ errors.

\subsection{GALFIT modelling for the SDSS Stripe~82 $g$- and $r$-band coadded images}
\label{sec:host_estimate_sdss}

\begin{figure*}
\center{
\includegraphics[clip, width=5.4in]{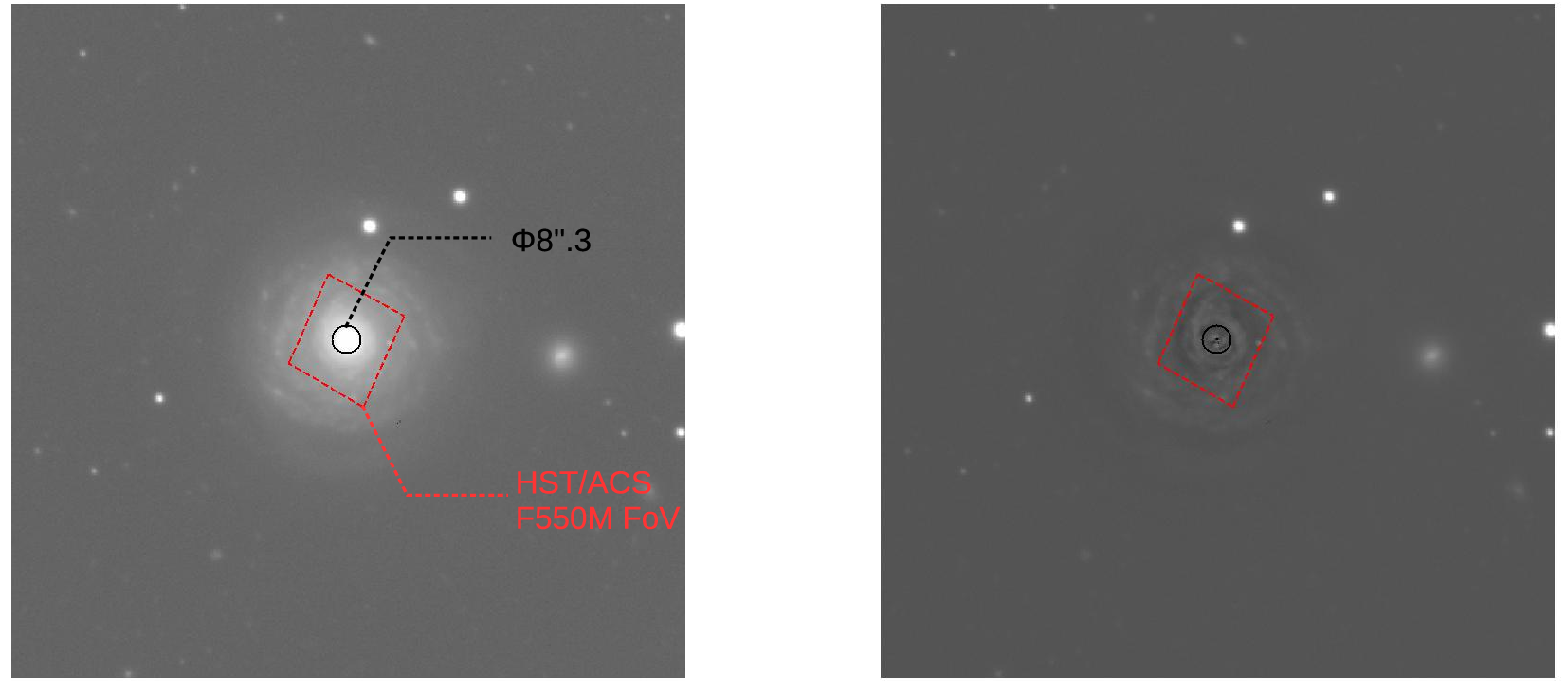}
}
 \caption{Left:the SDSS Stripe~82 $g$-band $3'.3 \times 3'.3$ cutout coadded image of Mrk~590. North is up and east is left. Right: the same image but with the best-fit GALFIT model of Mrk~590 subtracted. The red rectangular region indicates the field of view of the HSC/ACS F550M image (Fig.~\ref{fig:GALFIT}). The black circle is the $\phi 8''.3$ circular aperture.}
 \label{fig:GALFIT_SDSS}
\end{figure*}

\begin{table*}
\centering
\footnotesize
%\scriptsize
\caption{GALFIT modelling of the SDSS Stripe 82 coadded images of Mrk~590}
\label{tbl-sdss-galfit}
\begin{tabular}{c c c c c c c}
\hline
(1)   & (2)    &     (3)            &   (4)   & (5) & (6)   & (7)\\
Model & Filter & $m_{\text{ABmag}}$ & $R_{e}$ & $n$ & $b/a$ & Note\\
      &        &                    &  (kpc)  &     &       &     \\
\hline
$\rm{S\acute{e}rsic}$ fixed  & $g$ &  $17.3319$ & \dots & \dots &  \dots & PSF \\
 &  &  $16.4610$ & [$0.4026$] &  [$1.22$] & [0.6186] & Inner bulge \\
 &  &  $16.9919$ & [$0.6912$] &  [$0.59$] & [0.9713] & Bulge \\
 &  &  $15.4765$ & $1.9886$   &  [$1$]    & 0.8488   & Disk \\
 &  &  $13.6703$ & $9.9580$   &  [$1$]    & 0.9454   & Outer disk \\\hline
$\rm{S\acute{e}rsic}$ fixed  & $r$ &  $17.1009$ & \dots & \dots &  \dots & PSF \\
 &  &  $15.6840$ & [$0.4026$] &  [$1.22$] & [0.6186] & Inner bulge \\
 &  &  $16.0096$ & [$0.6912$] &  [$0.59$] & [0.9713] & Bulge \\
 &  &  $14.5052$ & $1.9498$   &  [$1$]    & 0.8708   & Disk \\
 &  &  $13.0160$ & $10.0464$  &  [$1$]    & 0.9375   & Outer disk \\\hline
\end{tabular}
\\
\begin{flushleft}
The parameters in square brackets are fixed. The zero AB magnitude is $30.000$~mag \citep{ann14}. The pixel scale is $0''.396$~pixel${}^{-1}$, and the angular scale of 0.495 kpc~$\text{arcsec}^{-1}$ is assumed.
\end{flushleft}
\end{table*}

We perform GALFIT modelling for the SDSS Stripe~82 coadded $g$- and $r$-band images of Mrk~590 \citep{ann14}.
The coadded images are sky-subtracted and have a uniform flux scale, such that 1 count corresponds to 30~ABmag.
The coadded image of Mrk~590 of each filter is constructed from 32 frames obtained prior to September 2005 during the SDSS Legacy Survey, and the effective gain of the coadded images are evaluated as 25.8 and 24.4 for the $g$ and $r$ bands, respectively.
The PSF FWHM values listed in the {\tt tsField} file are $1''.48$ and $1''.32$ for the $g$ and $r$ bands, respectively.
We also extract a model PSF for each coadded image, calculated by \cite{ann14} from the {\tt psField} file using the {\tt readAtlasImages-v5\_4\_11} program distributed on the SDSS website \citep[e.g.,][]{mee15}\footnote{\href{http://classic.sdss.org/dr7/products/images/read_psf.html}{http://classic.sdss.org/dr7/products/images/read\_psf.html}}.

The GALFIT modelling is applied to $3'.3 \times 3'.3$ cutout images centred on Mrk~590 (Fig.~\ref{fig:GALFIT_SDSS}).
Because the inner bulge and bulge components have small effective radii comparable to the PSF size and are only partially resolved, we fix the effective radii, $\rm{S\acute{e}rsic}$ indices, and axis ratios for the two components to the values constrained from the HST/ACS image decomposition (the ``$\rm{S\acute{e}rsic}$ fixed'' model in Table~\ref{tbl-galfit}).
In addition to the inner bulge, bulge, and disk components required to fit the HST/ACS image, we determine that the GALFIT modelling for the SDSS images requires an additional extended host galaxy component (outer disk component) with an effective radius of $R_{e} \sim 20'' \sim 10$~kpc.
The $\rm{S\acute{e}rsic}$ indices of the disk and outer disk components are fixed to 1.
Other than Mrk~590, four field stars and one $\rm{S\acute{e}rsic}$ galaxy near Mrk~590 are fitted simultaneously.

Table~\ref{tbl-sdss-galfit} shows the best-fitting results for Mrk~590 imaged on SDSS coadded images.
Then, by performing $\phi 8''.3$ circular aperture photometry for PSF-subtracted SDSS coadded images, the host galaxy fluxes within the $\phi 8''.3$ circular aperture are evaluated as $3.373^{+0.029}_{-0.029}$~mJy and $7.538^{+0.047}_{-0.047}$~mJy for the $g$- and $r$-band images, respectively.

%%%%%%%%%%%%%%%%%%%%%%%%%%%%%%%%%%%%%%%%%%%%%%%%%%


\bibliography{Mrk590_reverberation_mapping_MNRAS_version2.bbl}

%-------------------------------------------------------------
%                 A figure as large as the width of the column
%-------------------------------------------------------------
%   \begin{figure}
%   \centering
%   \includegraphics[width=\hsize]{empty.eps}
%      \caption{Vibrational stability equation of state
%               $S_{\mathrm{vib}}(\lg e, \lg \rho)$.
%               $>0$ means vibrational stability.
%              }
%         \label{FigVibStab}
%   \end{figure}

% Don't change these lines
\bsp	% typesetting comment
\label{lastpage}
\end{document}